\documentstyle[prl,aps,multicol,epsf]{revtex} 
\input epsf 
\newcommand{\E}{{\rm e}} 
\newcommand{\D}{{\rm d}} 
\newcommand{\I}{{\rm i}} 
\newcommand{\Tr}{{\rm Tr}} 
\renewcommand{\Re}{{\rm Re}}

\firstfigfalse

\begin{document} 
\title{Coherent transport in disordered metals out of equilibrium} 
\author{P. Schwab$^{(1)}$ and R. Raimondi$^{(2)}$} 
\address{ 
 $^{(1)}$Institut f\"ur Physik, Universit\"at Augsburg, D-86135 Augsburg \\ 
 $^{(2)}$NEST-INFM e Dipartimento di Fisica "E. Amaldi", 
 Universit\`a di Roma3, 
	Via della Vasca Navale 84, 00146 Roma, Italy
} 
\date{\today} 
\maketitle 
\begin{abstract} 
We derive a formula for the quantum corrections to the electrical current for
a metal out of equilibrium. In the limit of linear current-voltage characteristics
our  formula reproduces the well known 
Altshuler-Aronov correction to the conductivity of a disordered metal.
The current formula is obtained by a direct diagrammatic 
approach, and is shown to agree with what is obtained within  the Keldysh formulation
of the non-linear sigma model.
As an application we calculate the current   of a mesoscopic wire. 
We find a current-voltage characteristics that scales with $eV/kT$, and calculate
the different scaling curves for a wire in the hot-electron regime  and in the
regime of full non-equilibrium.
\end{abstract} 
 
\begin{multicols}{2} 
\section{Introduction} 
Quantum interference effects in disordered metals have been the subject of 
intensive investigation for over twenty years.
For general reviews see\cite{lee85,altshuler85,bergmann84}.
The interference of the scattered electrical waves in the presence of 
a random potential leads 
to corrections\cite{gorkov79,abrahams80,altshuler80}
to the semi-classical formula of the electrical conductivity known from the
Drude-Boltzmann theory. 
The physical implications of  these quantum corrections have been extensively
discussed in the literature mostly for the equilibrium properties,
for which experimental data were available, 
although a number of non-linear electric field effects have been predicted in the past
\cite{altshuler85,larkin86,falko89,kravtsov93}.

In contrast, non-equilibrium electrical transport has received 
considerable attention in the field of mesoscopic physics.
Examples are  the transport in quantum dots\cite{quantumdot},
or the shot noise in mesoscopic conductors\cite{blanter00}. 
In this situation however the majority of the studied phenomena did not 
involve interference effects.

Our  interest in the non-equilibrium properties of interference phenomena 
originated  by  the
suggestion\cite{altshuler98} that the non-equilibrium electric noise could be the origin for the
low temperature saturation of the weak localization  dephasing time observed in
disordered films and wires\cite{mohanty97}.
Mohanty et al.\cite{mohanty98} pointed out that in the samples with the strongest dephasing
rate also the  interaction correction (Altshuler-Aronov)
to the conductivity saturates at low temperature.
This suggests that  
also the Altshuler-Aronov correction should be affected by the non-equilibrium noise.
Similar speculations concerning
an electric field effect on the Altshuler-Aronov correction have also been made
earlier by different authors in different contexts\cite{bergmann90,giordano91}.

This motivated us to study the interaction correction in the presence of an external field.
In analogy to the analysis  performed in the literature 
for weak localization\cite{altshuler85},
we calculated the interaction correction to the current in the presence of 
a time dependent vector potential, assuming a thermal distribution function
\cite{raimondi99}. 
Indeed we verified the above-mentioned speculations since we 
found an electric field effect. However, contrary to what is known in the
case of weak localization, where the strongest effect occurs when the period
of the AC field is of the order of the dephasing time,  we found
a suppression of the interaction correction even by static electric fields.
These findings have however raised the issues of the possibility of experimentally
observing the effect and of its physical interpretation.
Both these questions will be addressed in this paper.

Let us comment first  the problem concerning the interpretation of the effect.
In Ref.\cite{leadbeater00} we demonstrated
that the non-linear field effect can be understood in terms of dephasing by calculating
the phase shifts of the relevant classical paths in the presence of a time dependent
vector potential.
On the other hand, a static electric field can also be described in terms of a 
static scalar potential, and it is clear that a static scalar potential does not
lead to dephasing.
Therefore we think it is of interest to present a version of our theory in the scalar gauge.
In this paper we will explicitly show the gauge invariance of our previous results.

The second problem concerns the scale of the effect and its experimental observability.
We found in Ref.\cite{raimondi99} that 
the temperature-dependent Altshuler-Aronov correction to the conductivity saturates when  
the voltage drop on the thermal length ($L_T =\sqrt{D/T}$) is comparable to the temperature 
$e E L_T \sim k T$. From this condition one can estimate
the strength of the microwave field  that is necessary to explain the  
saturation of the Altshuler-Aronov correction observed in the experimental data of
Ref.\cite{mohanty98}. In so doing one arrives at a microwave field value 
that is more than an order of magnitude  larger than
the optimistic estimate given in Ref.\cite{altshuler98} 
to explain the saturation of the weak localization time. 
Furthermore it is rather unlikely that the condition $ e E L_T \sim k T$ can
be reached at low temperature, since strong heating is already assumed
to set in when the voltage drop over the electron-phonon length is of order
of the temperature.
Strong heating on the other hand is not observed in Ref.\cite{mohanty98}.
Similar problems arise also in the attempt to explain the experimental data of 
Refs.\cite{bergmann90,giordano91}.

Despite the above mentioned problems,  we think we cannot rule out non-equilibrium noise as a reason 
for the observed saturations. In fact,  even in the absence of strong heating,
the distribution function may deviate from the equilibrium form and  affect the interaction correction
to the conductivity and possibly lead to saturation at considerably weaker electric fields.
A theory which is valid even out of equilibrium will be developed in this paper.

In contrast to our previous work\cite{raimondi99} we will  
avoid to guess the distribution function
which could be relevant for the experiments of Refs.\cite{mohanty97,mohanty98}. 
Instead we will calculate the interaction correction in a more controlled situation.
Nowadays it is, indeed,  possible to create non-equilibrium in a controlled way by, for instance,
attaching a short mesoscopic wire to large metallic reservoirs 
(see e.g.\cite{pothier97}). 
In the absence of inelastic scattering processes
the distribution function in the wire is a linear superposition of the distribution functions
of the leads.
The interaction correction in such a situation has also been considered in the recent paper 
by Gutmann and Gefen\cite{gutmann00}.
We verify their result that the $I-V$ characteristics scales as $eV/kT$.
Going beyond, we calculate the $I-V$ characteristics explicitly and we will compare
quantitatively the wire in non-equilibrium with the wire in the hot electron regime.
In addition we will also discuss the interaction correction in the spin triplet channels.

Our paper is organized as follows.
In the next section we introduce the basic quantities and recall
the main results of the Drude-Boltzmann theory within the Keldysh formalism.
In section III we consider the quantum corrections to the conductivity
within the Keldysh diagrammatic approach. We derive, in particular,
 an expression for the
current in the presence of an external electric field.
In section IV we discuss the gauge invariance of the theory, 
while in section V
we present a specific application: a mesoscopic wire.
Finally in section VI we give our conclusions.
In the appendices we outline how to obtain the same results using 
the Keldysh formulation of the non-linear sigma model and 
we extend the calculations in order to include also the spin effects.

\section{Basic definitions and the Drude-Boltzmann theory}
In this section we will recall some basic relations of the quasi-classical approximation in its 
non-equilibrium (Keldysh) formulation\cite{keldysh64}.
Our notation will mainly follow Ref.\cite{rammer86}.
We will write down the equation of motion for the Green functions 
in the presence of impurity scattering in the case when quantum 
interference is completely neglected.
The Green functions have the matrix structure
\begin{equation} \label{eq1}
\hat G = \left( \begin{array}{cc} 
G^R & G^K \\
0   & G^A 
  \end{array} \right)
,\end{equation}
with
\begin{eqnarray} 
G^R(x,x') & =  &-\I \Theta( t -t') 
\left( \langle \Psi(x) \Psi^\dagger(x')
\! + \!
\Psi^\dagger(x')\Psi(x)\rangle \right) \\
G^A(x,x') & =  &+\I \Theta( t' -t)\left( 
\langle \Psi(x) \Psi^\dagger(x') 
\! + \!
\Psi^\dagger(x')\Psi(x)\rangle  \right) \\
G^K(x,x') & =  &-\I \left( 
\langle \Psi(x) \Psi^\dagger(x') 
\! -  \!
\Psi^\dagger(x')\Psi(x)\rangle \right) 
,\end{eqnarray}
where $\Psi$ and $\Psi^\dagger$ are fermion operators and $x= ({\bf x}, t)$.
In equilibrium the Keldysh component of the Green function
is expressed in terms of   the retarded and advanced components by
$G^K_\epsilon = [1-2 f(\epsilon)] (G^R_\epsilon-  G^A_\epsilon )$,
where $f(\epsilon)$ is the Fermi function.
The Keldysh component out of equilibrium will be discussed later. 

\begin{figure}
\noindent
\begin{minipage}[t]{0.98\linewidth} 
\hspace{0.5cm}{\epsfxsize=3.5cm\epsfbox{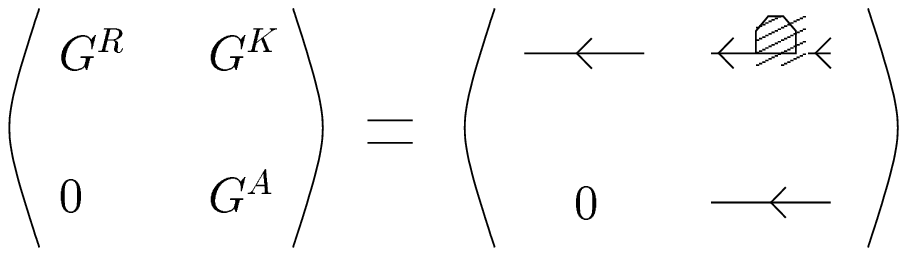}} 
\vspace{1cm}
\caption{
Graphical representation of the Green function; the shaded box in $G^K$ represents
the Keldysh component of the self-energy, i.e.,  basically the distribution 
function.}
\label{fig1}
\end{minipage}
\end{figure}

The Green function solves the differential equation
\begin{eqnarray} 
\left( \I{\partial \over \partial t} +{1\over 2 m}
(\nabla + \I e {\bf A})^2 + e \phi + \mu  \right) 
\hat G({\bf x},t ;{\bf x'}, t' ) && \nonumber\\
 -\int \D t_1 \D {\bf x}_1 \hat \Sigma({\bf x},t;{\bf x}_1,t_1 )
\hat G({\bf x}_1,t_1 ;{\bf x'}, t' ) \nonumber\\
= \delta( {\bf x} -{\bf x'}) 
\delta( t-t')
,&&\end{eqnarray}
where 
and $\phi$, ${\bf A}$ are the scalar and vector potential.
Since the self-energy $\hat \Sigma$ has the same triangular matrix 
structure as the 
Green function,  one can invert the inverse Green function
$\hat G^{-1}$ and finds for the Keldysh component
the relation
\begin{equation} \label{eq9}
G^K = G^R \Sigma^K G^A
.\end{equation}
For a graphical representation see Fig.\ref{fig1}.
We then introduce the $\xi$-integrated (quasi-classical) 
Green function
\begin{equation} \label{eq10}
\hat g_{tt'} ( {\bf {\hat p} }, {\bf R} )
= {\I \over \pi} \int \D \xi
\D {\bf r} \E^{- \I {\bf p} \cdot {\bf r} }
\hat G\left( {\bf R}+{ {\bf r} \over 2}, t; 
        {\bf R}-{ {\bf r} \over 2}, t'\right)
,\end{equation}
where $\xi = {\bf p}^2/2m - \mu $ and ${\bf {\hat p}}$ is a unit vector
along the momentum.
The Green function in the energy domain is
\begin{equation} \label{eq14}
\hat g_{\epsilon \epsilon'} ({\bf {\hat p} }, {\bf x})
= \int \D t \D t' \E^{\I \epsilon t - \I \epsilon' t' }
\hat g_{tt'} ({\bf {\hat p}}, {\bf x} ).
\end{equation}
We will keep the notation of small $g$ for the $\xi$-integrated Green
functions and capital $G$ for the not integrated Green functions 
all over this paper.
When approximating the density of states as an energy independent constant, the
$\xi$-integration is related to an integration over the momentum ${\bf p}$ 
according to
\begin{equation} \label{eq11}
\int {\D^3 p \over (2\pi)^3 } \to
N_0 \int \D \xi \int {\D \hat p \over 4 \pi}
.\end{equation}

We will now recall some relations that are specific for impurity scattering.
By treating the impurity scattering within the self-consistent Born approximation
and assuming a Gaussian, $\delta$-correlated impurity potential
with
\begin{equation}
\langle U({\bf x}) U({\bf x}') \rangle =
{1\over 2 \pi N_0 \tau } \delta( {\bf x} - {\bf x'} ),
\end{equation}
the electron self-energy is local in space and is given by
\begin{eqnarray}
\hat \Sigma^{\rm imp}( {\bf x},t ; {\bf x}' ,t' )& = &  
{1\over 2 \pi N_0 \tau}   \hat G( {\bf x},t; {\bf x}, t' ) \delta({\bf x} - {\bf x}')\\
&=& \hat \Sigma^{\rm imp}_{tt'}({\bf x}) \delta({\bf x} - { \bf x'} )
.\end{eqnarray}
Notice that this equation has to be solved self-consistently for all
the components of the Green function.
Using the above definition,  one observes that the impurity self-energy
is related to the $s$-wave part of the quasi-classical function,
\begin{equation} \label{eq12}
\hat \Sigma^{\rm imp}_{tt'}({\bf x})  = -{\I \over 2 \tau}
\int {\D \hat p \over 4 \pi} \hat g_{tt'} ({\bf {\hat p}} ,{\bf x}) 
.\end{equation}

The distribution function out of equilibrium is found by solving 
the appropriate kinetic equation.
Here we derive the kinetic equation for $g^K_{\epsilon \epsilon'} $ 
from Eqs.(\ref{eq9}) and (\ref{eq10}). For
simplicity we neglect external fields for the time being.
Under these conditions the retarded and advanced Green functions are
\begin{equation}
G^{R(A)}({\bf p}, \epsilon) = {1\over \epsilon - \xi \pm \I/2\tau }.
\end{equation}
Near the Fermi energy ($\epsilon, \epsilon' \ll \epsilon_F$) and
for small momenta ($ q \ll p_F$) one finds
\begin{eqnarray} \label{eq16}
g^K_{\epsilon \epsilon'} ({\bf {\hat p}}, {\bf q} ) & = &
{\I \over \pi}\int {\D^3 p \over (2\pi)^3}
G^R(\epsilon, {\bf p} + {\bf q} /2 ) \nonumber\\
&&\times
\Sigma^K_{\epsilon \epsilon'}({\bf q} )
G^A(\epsilon', {\bf p } - {\bf q}/2) \\
\label{eq17}
& \approx  & {\I \over \tau} {1\over \epsilon  - \epsilon' +\I /\tau -
v_F\hat {\bf p} \cdot {\bf q} }
\int \!{\D \hat  p \over 4 \pi } g^K_{\epsilon \epsilon'} 
({\bf  {\hat p}} , {\bf q}).
\end{eqnarray}
The equation above reproduces the well-known  
kinetic equation for impurity scattering
\begin{eqnarray}
\lefteqn{ \left( {\partial \over \partial t} +
                 {\partial \over \partial t'}+
v_F {\bf {\hat p}} \cdot {\bf \nabla}\right) 
g^K_{tt'} ({\bf {\hat p}}, {\bf x} )
=} && 
\nonumber\\
& &\frac{1}{\tau}\left( g^K_{tt'} ({\bf {\hat p}}, {\bf x} )- 
\int{\D \hat p \over 4 \pi } g^K_{tt'} ({\bf {\hat p}}, {\bf x} ) \right). 
\end{eqnarray}
In this work we will restrict to the case, where energies and
momenta are restricted even more, namely
$\epsilon \tau, \epsilon' \tau  , q v_F \tau \ll 1 $.
By expanding (\ref{eq17}) for small energy and momentum and taking the
angular average,  one finds the diffusive equation
\begin{equation} 
\left( 
{\partial \over \partial t} +
{\partial \over \partial t'} - 
D {\partial^2 \over \partial {\bf x}^2  } \right)
\int{\D \hat p \over 4 \pi } g^K_{tt'} ({\bf {\hat p}}, {\bf x} ) = 0 
,\end{equation} 
where the diffusion constant is $D= v_F^2\tau /3 $.
Notice that this equation is solved by any function
$g^K_{tt'}({\bf p}, {\bf x})$ which is independent of position ${\bf x}$
and which depends on time differences $(t-t')$ only.
This reflects the fact 
that any distribution function is allowed for noninteracting
electrons.

The charge density and current density are related to the Keldysh component of
the Green function,
\begin{eqnarray} \label{eq19}
\rho({\bf x}, t)   \! &=& \!
\I e G^K({\bf x}, t ; {\bf x}, t) \\
\label{eq20}
{\bf j}({\bf x},t) \! &=& \!
{e\over {2m}}
\left[ \nabla_{\bf x} \! - \! \nabla_{\bf x'} \! + \! 2\I e A({\bf x},t ) \right]
G^K({\bf x}, t; {\bf x'}, t)|_{{\bf x'}={\bf x}}.
\end{eqnarray}
In terms of the quasi-classical Green functions, the charge and current read \cite{rammer86}
\begin{eqnarray} \label{eq21}
\rho({\bf x},t) &=& 2e N_0 \left(
{\pi \over 2} \int { \D\hat p \over 4 \pi} g^K_{tt} ({\bf {\hat p}}, {\bf x}) 
- e \phi({\bf x}, t) \right) \\
\label{eq22}
{\bf j}({\bf x}, t) &=& e \pi N_0 \int { \D \hat p \over 4\pi}
 v_F {\bf \hat p}  g^K_{tt} ({\bf {\hat p}}, {\bf x} ).
\end{eqnarray}

It is useful to consider the current density in
the presence of an electric field 
${\bf E}({\bf x})= -\nabla \phi({\bf x})$.
By replacing $G^K$ in (\ref{eq20}) with $G^R \Sigma^K G^A$ we express the current density as
\begin{eqnarray} \label{eq23}
{\bf j}({\bf q},\omega )  &=& \I e  
\int {\D \epsilon \over 2 \pi }
\int {\D^3 p \over (2\pi)^3} 
{ {\bf p} \over m }
G^R\left( \epsilon+{\omega\over 2}, {\bf p} +{ {\bf q} \over 2} \right)
\nonumber\\
&&\times
\Sigma^K_{\epsilon+\omega/2, \epsilon-\omega/2}({\bf q} ) 
G^A\left(
\epsilon-{\omega\over 2}, {\bf p} -{ {\bf q}\over 2} \right) 
,\end{eqnarray}
from which we obtain 
\begin{eqnarray}
\label{eq24}
{\bf j}({\bf x},t ) & = & 
-e \pi D  N_0  \nabla 
\int{\D \hat p \over 4 \pi } 
g^K_{t t} ({\bf {\hat p}}, {\bf x})
\\
&=&
-D \nabla \rho({\bf x},t) 
+ 2 e^2 D N_0 {\bf E}({\bf x},t ).
\end{eqnarray}
Within the adopted approximations , i.e., a constant density of states and 
a uniform diffusion coefficient,
one observes that the current is a linear function of the electric field as
long as the 
charge density $\rho({\bf x})$ stays uniform. 

We close this section by  commenting on Eq.(\ref{eq21}).
A scalar field $\phi({\bf x}, t)$ shifts the entire
Fermi surface, i.e., it  affects the Green function at all energies.
This is lost 
in the  naive substitution of  Eq.(\ref{eq11}), when the $\xi$-integration 
is extended to $\pm \infty$.
The second term in eq.(\ref{eq21}) is obtained by taking into account these high energy 
terms correctly.
The equilibrium response to a static field  is, for instance,
fully given by this second contribution. 
For this reason the second term is often referred to as the ``static contribution''
to the response, whereas the first term is referred to as the 
``dynamic contribution''.
%
%
%
\section{Quantum correction to the current}
Quantum interference gives rise to corrections to the semi-classical
expression of the electrical conductivity of a metal.
The so-called quantum corrections to the average conductivity 
are the weak localization
correction (WL), the interaction correction in the particle-hole channel (EEI), and
 the interaction correction in the Cooper channel (EEIC). 
In this paper we will
concentrate on the interaction 
correction in the particle-hole channel.
For non-linear effects in WL we refer to the literature \cite{altshuler85,bergmann84}.
Interactions in the Cooper channel will not be considered. This is justified
for non-super-conducting materials since in that situation
the relevant interaction  parameter scales 
downwards under the renormalization group.

\subsection{Ladder diagrams}
Before calculating the quantum corrections we introduce the
ladder diagrams of repeated impurity scattering which will appear at
various places in the diagrammatic approach. 
Technically speaking these ladder diagrams appear when averaging a product
of a retarded and an advanced Green function.
Here we  briefly recall how to derive  the expressions for the ladder
in the absence of external fields and
without spin effects. The inclusion of external fields and spin structure is
straightforward and one may refers to  the reviews
on the subject like Ref.\cite{altshuler85}.

\begin{figure}
\begin{minipage}[t]{0.98\linewidth}
\hspace{0.5cm}{\epsfxsize=7cm\epsfbox{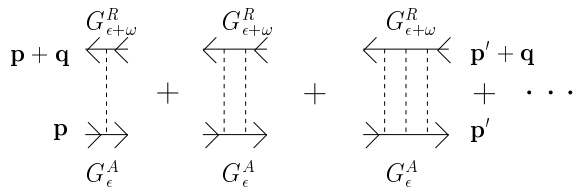}} 
\vspace{1cm}
\caption{Graphical definition of the diffuson (particle-hole ladder).
$G^R$ and $G^A$ are the retarded and advanced Green functions, which in the
general case can depend on external electromagnetic fields.}
\label{fig3}
\end{minipage}
\end{figure}

\begin{figure}

\noindent
\begin{minipage}[t]{0.98\linewidth}
\hspace{0.5cm}{\epsfxsize=7cm\epsfbox{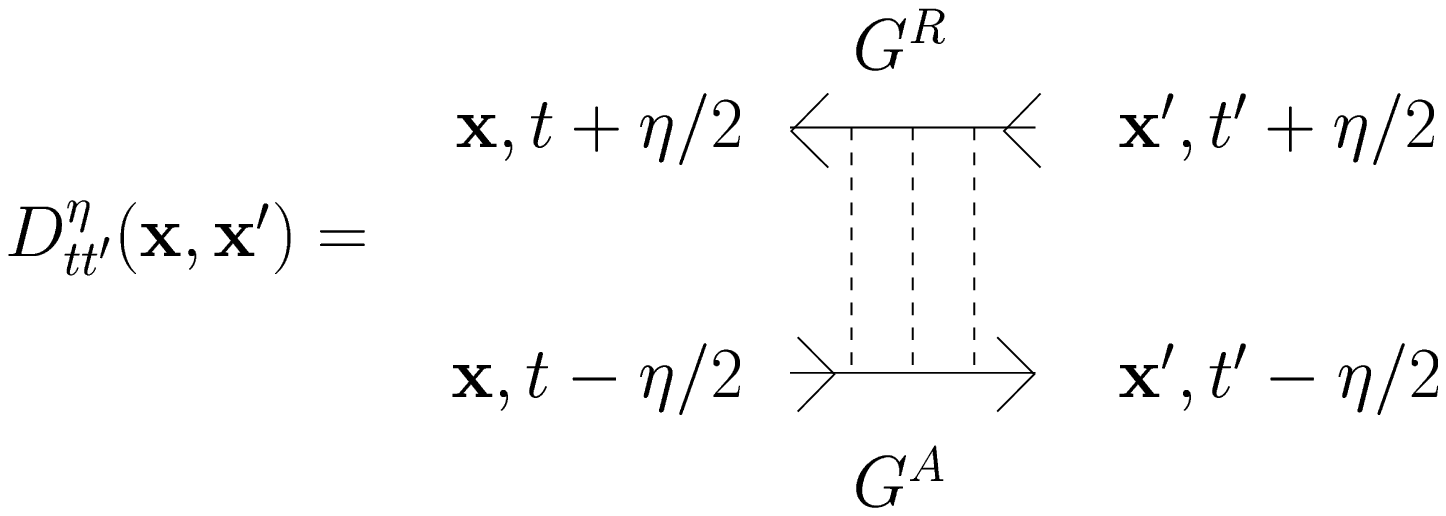}} 
\vspace{1cm}
\caption{
The diffuson in the space/time domain.
}
\label{fig5}
\end{minipage}
\end{figure}

The diffuson $D({\bf q}, \omega)$ 
or particle-hole ladder is found by summing the sequence
of diagrams shown in Fig.\ref{fig3}: 
\begin{equation}
D({\bf q}, \omega) = 1 + \eta^{RA} +(\eta^{RA})^2 + \cdots = {1\over 1-\eta^{RA}}
\end{equation}
with
\begin{eqnarray}
\eta^{RA} & =&{1\over 2 \pi N_0 \tau} 
 \int{\D^3 p \over (2\pi)^3 }G^R(\epsilon+\omega, {\bf p} + {\bf q} )
G^A(\epsilon, {\bf p}) \\
 &\approx & 1  - \tau ( -\I \omega + Dq^2 )
\end{eqnarray}
where we have used the condition that $\omega \tau \ll 1$ and $v_Fq\tau\ll 1$ 
so that the diffuson reads
\begin{equation}
D({\bf q}, \omega )= {1\over \tau }{1\over -\I \omega + Dq^2 }.
\end{equation}
For completeness we give the expression in
the presence of external electromagnetic fields. In this case
it is convenient to go a real space representation where  the
diffuson is defined by the equation
\begin{eqnarray}
\label{diffusonequation}
\left\{ {\partial \over \partial t} - D (\nabla_{\bf x}
+\I e {\bf A}_D )^2 - 
 \I e \phi_D 
\right\} D^{\eta}_{tt'}({\bf x},{\bf x}') && \nonumber\\
\label{eq31}
= {1 \over \tau} \delta({\bf x} - {\bf x}' ) \delta( t-t' )
,\end{eqnarray}
with 
${\bf A}_D = {\bf A}({\bf x},t+\eta/2)
- {\bf A}({\bf x}, t-\eta/2 )$ and 
$\phi_D = \phi({\bf x}, t+\eta/2)- \phi({\bf x}, t-\eta/2)$.
In these equations $t$ is the center-of-mass time, and $\eta$ is the relative
time, and are defined in Fig.\ref{fig5}.
Notice that the external field drops from the equation for the diffuson when
the relative time $\eta$ equals zero.
\subsection{Interaction correction to the current}
\begin{figure}
\noindent
\begin{minipage}[t]{0.98\linewidth}
\hspace{0.5cm}{\epsfxsize=4cm\epsfbox{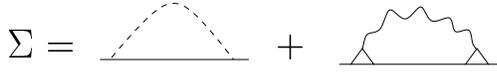}} 
\vspace{1cm}
\caption{The self-energy containing both disorder and interaction.}
\label{fig7a}
\end{minipage}
\end{figure}

\begin{figure}
\noindent
\begin{minipage}[t]{0.98\linewidth}
\hspace{2.0cm}{\epsfxsize=2.0cm\epsfbox{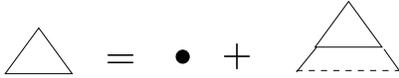}} 
\vspace{1cm}
\caption{Dressing of the interaction vertex with impurity lines.}
\label{fig7b}
\end{minipage}
\end{figure}

We are now ready to allow for electron-electron interactions.
Interactions will enter the kinetic equation and 
determine the form of the distribution function.
We assume that the distribution function 
has been determined self-consistently via
the kinetic equation with the inclusion of the interaction.
We will concentrate then on the calculation of the interaction corrections to the current density.
To do so we need the expression for the Keldysh Green function in the presence
of interactions.
Following Refs.\cite{altshuler78,rammer86} we start with the self-energy
\begin{equation}
\Sigma = \Sigma^{\rm imp} + \Sigma^V
\end{equation}
where $\Sigma^{\rm imp}$ is the previously defined impurity self-energy 
and
\begin{eqnarray}
\Sigma^V_{ij}(x,x') &= & \I \sum_{i'j' kk'}\int \D x_2 \D x_3 \D x_4 \D x_5
\Gamma^k_{ii'}(x_5; x, x_3 ) \cr
&\times &V^{kk'}(x_5, x_4)
G_{i'j'}(x_3, x_2) \tilde \Gamma^{k'}_{j'j}(x_4; x_2,x').
\end{eqnarray}
The vertex functions are given by
\begin{eqnarray}
\Gamma^k_{ij}(x;x_1, x_2) &= & \gamma^k_{ij} + {1\over 2\pi N_0 \tau}
\sum_{i'j'} \int \D x_1' \D x_2' 
G_{ii'}(x_1, x_1')\cr
&\times &\Gamma^k_{i'j'}(x; x_1', x_2')G_{j'j}(x_2', x_2)
.\end{eqnarray}
An analogous equation is valid for $\tilde \Gamma$.
We recall that in the Keldysh triangular representation the ``absorption'' and
``emission'' vertices differ.
A diagrammatic representation of both the self-energy and vertex equations is shown in
Figs.\ref{fig7a} and \ref{fig7b}.
The indices $i,j, \dots$ denote matrix indices in Keldysh space.
The bare vertices $\gamma$, $\tilde\gamma$ are  local in space and time. 
The structure in Keldysh space is 
$\gamma^1_{ij} = \tilde \gamma^2_{ij} = \delta_{ij}/\sqrt{2}$
and 
$\gamma^2_{ij} = \tilde \gamma^1_{ij} = \sigma^x_{ij}/ \sqrt{2}$.
From $\Sigma^{\rm imp}+\Sigma^V$ one may derive a kinetic equation 
for the system with disorder and interaction.
The general expressions were already given in the seminal paper, 
Ref.\cite{altshuler78}. 
Explicit 
expressions for the various components of $\Gamma$ and $\tilde \Gamma$ 
in terms of integrals like $\eta^{RA}$ are also
given in the appendix of Ref.\cite{rammer86}.
Fully {\em evaluating} the expressions in this or that limit remains still
to be done. 
Even in thermal equilibrium we are not aware of any full self-consistent calculation.

\begin{figure}

\noindent
\begin{minipage}[t]{0.98\linewidth}
\hspace{0.5cm}{\epsfxsize=7.5cm\epsfbox{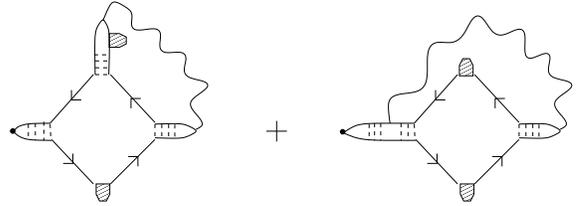}} \\ 
\vspace{0.5cm}
\caption{
Interaction correction to the current; diagrams of the type shown here
correspond to $\delta {\bf j}_a $ and may be related to the gradient
of the density.
}
\label{fig8}
\end{minipage}
\end{figure}

In the following we will take into account only 
the non-interacting self-energy
self-consistently,
and restrict ourselves to the perturbation theory for the interacting part of the problem.
The change in $G^K$ due to the interaction may be written as 
\begin{equation}
\delta G^K =G^R \delta \Sigma^R G^K +  G^R \delta \Sigma^K G^A 
           + G^K \delta \Sigma^A G^A
,\end{equation}
where $\delta \Sigma$ is a sum of the interaction self-energy plus the 
interaction-induced change in the impurity self-energy
\begin{equation}
\delta \Sigma = \delta \Sigma^{\rm imp} + \Sigma^V
.\end{equation}
Among the many contributions to $\delta \Sigma$ we start with the Keldysh
component of $\delta \Sigma^{\rm imp}$. We denote the corresponding correction to the current
as $\delta {\bf j}_a$, which we determine as
\begin{equation}
\delta {\bf j}_a({\bf x}, t) =- \I e D 2 \pi N_0 \tau \nabla
\delta \Sigma^{{\rm imp},K}
.\end{equation}
Apparently $\delta {\bf j}_a$ is related to the correction
to the charge density and may be written as
\begin{eqnarray}
\delta {\bf j}_a({\bf x}, t) &=& -e \pi D N_0 \nabla
\int {\D \hat  p \over 4\pi} \delta g^K_{tt}({\bf p},{\bf x}) \\ 
\label{eqj_a}            &=& -D \nabla \delta \rho({\bf x} ).
\end{eqnarray}
Some typical diagrams contributing to $\delta {\bf j}_a$ are shown in Fig.\ref{fig8}.
\begin{figure}
The calculation of the correction to the current from all the other contributions,
which we will denote by $\delta {\bf j}_b$,
simplifies due to the following observation:
the majority of the components of the renormalized vertices $\Gamma$, $\tilde \Gamma$ 
are strongly enhanced over
the bare value due to the presence of  a diffusive type of vertex correction.
The explicit calculation \cite{rammer86} shows, however, 
that the diagonal parts of $\Gamma^1_{ij}$ and
$\tilde \Gamma^2_{ij}$ are not renormalized. In the leading order of a gradient
expansion we can therefore neglect all the terms involving these vertices.
As a result the expression for the various self-energy components are given by  the combinations\cite{rammer86}
\begin{eqnarray}
\Sigma^{V,R} &:& \Gamma^1_{12} G_{22} \tilde \Gamma^1_{21}V^R \\
\Sigma^{V,A} &:& \Gamma^2_{21} G_{11} \tilde \Gamma^2_{12}V^A \\
\Sigma^{V,K} &:& \Gamma^1_{12} G_{22} \tilde \Gamma^1_{22} V^R + 
                 \Gamma^2_{11} G_{11} \tilde \Gamma^2_{12} V^A 
\end{eqnarray}
and
\begin{eqnarray}
\delta \Sigma^{{\rm imp},R} &:& G^R \Sigma^{V,R} G^R/(2\pi N_0 \tau) \\
\delta \Sigma^{{\rm imp},A} &:& G^A \Sigma^{V,A} G^A/(2\pi N_0 \tau). 
\end{eqnarray}
A diagrammatic representation of the correction to the current from
$\Sigma^{V,R}$ and $\Sigma^{V,A}$ is shown in Fig.\ref{fig9}.

\noindent
\begin{minipage}[t]{0.98\linewidth}
\hspace{0.5cm}{\epsfxsize=7.5cm\epsfbox{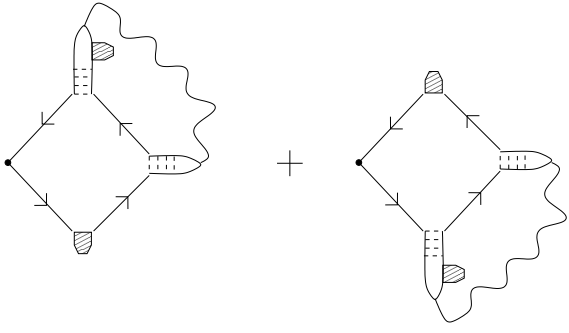}} \\ 
\vspace{0.5cm}
\caption{Interaction correction to the current; these diagrams contribute to $\delta j_b$.}
\label{fig9}
\end{minipage}
\end{figure}
Going through the algebra one convinces one-self that
the contributions from $\Sigma^{V,K}$ and $\delta \Sigma^{\rm imp}$ may be combined as
\begin{eqnarray}
\Sigma^{V,R} +\delta\Sigma^{{\rm imp},R}+\Sigma^{V,K}_a \\
\Sigma^{V,A} +\delta\Sigma^{{\rm imp},A}+\Sigma^{V,K}_b 
\end{eqnarray}
where $\Sigma^{V,K}_a$ and $\Sigma^{V,K}_b$ refer to the two terms entering the expression
for $\Sigma^{V,K}$. These combinations of terms 
are taken into account in the diagrams of Fig.\ref{fig9}
by completing the ``Hikami box'' as shown in Fig.\ref{fig9a}.
\begin{figure}
\noindent
\begin{minipage}[t]{0.98\linewidth}
\hspace{0.5cm}{\epsfxsize=7.5cm\epsfbox{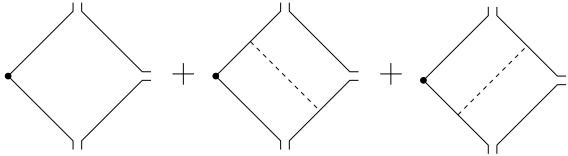}} \\ 
\caption{The three diagrams constituting the Hikami box; within
the here-applied formalism the diagrams are generated from the self-energies
$\delta\Sigma^{\rm imp}+ \Sigma^V$.}
\label{fig9a}
\end{minipage}
\end{figure}
We can now proceed to the  explicit calculation of the correction to the current.
Let us start with the evaluation of the Hikami box.
The first of the diagrams of the Hikami box
is shown in more detail in Fig.\ref{fig10}.
\begin{figure}
\noindent
\begin{minipage}[t]{0.98\linewidth}
\hspace{0.5cm}{\epsfxsize=7.5cm\epsfbox{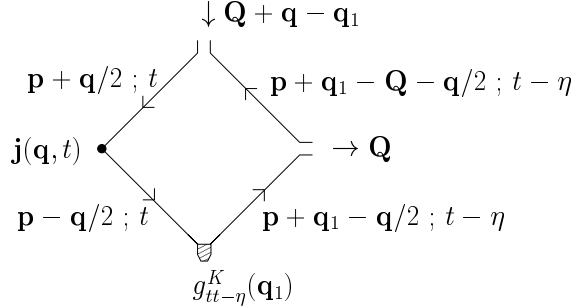}} \\ 
\vspace{1cm}
\caption{
Labeling of momenta and times in the Hikami box;
within the here-applied approximations 
the retarded and advanced Green functions are local in time.
}
\label{fig10}
\end{minipage}
\end{figure}
The evaluation of the diagram amounts to the integration over the ``fast'' momentum 
entering the electron Green function and is accomplished by
\begin{eqnarray}
&&\int {\D^3 p \over (2\pi)^3} { {\bf p }\over m}
G^R({\bf p} +{\bf q}/2 ) 
G^A({\bf p} -{\bf q}/2 ) \nonumber\\
&&\times G^R({\bf p} -{\bf q}/2+{\bf q}_1 ) 
G^A({\bf p} -{\bf q}/2+{\bf q}_1 -{\bf Q} ) 
.\end{eqnarray}
The integration is performed under the assumption that the three momenta
${\bf q}$, ${\bf q}_1$, and ${\bf Q}$ are small compared to $1/l$ and
that the energy in all the four Green functions is small compared
to $1/\tau$.
After expanding the Green functions to first order in
${\bf q}$, ${\bf q}_1$, $ \bf Q$,
the integration over 
${\bf p}$ gives
$(- 6\pi \I N_0 \tau^2)  D({\bf Q} +{\bf q })$.
Evaluating the other two diagrams of Fig.\ref{fig9a} analogously and summing up the three of them 
one arrives at
$(- 4\pi \I N_0 \tau^3 ) D {\bf Q}$.

\begin{figure}
\noindent
\begin{minipage}[t]{0.98\linewidth}
\hspace{2.5cm}{\epsfxsize=2.5cm\epsfbox{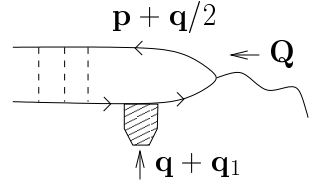}} \\ 
\caption{The interaction vertex; compare Fig.\ref{fig9}. }
\label{fig11}
\end{minipage}
\end{figure}

After completing the evaluation of the Hikami box, we now address our attention to
the vertex functions $\Gamma$ and $\tilde\Gamma$.
From Ref.\cite{rammer86} we borrow the relevant expressions as
\begin{eqnarray}
\Gamma^1_{12} &=& {1\over \sqrt{2} } (1-\eta^{RA} )^{-1}( \eta^{RK}+\eta^{KA}) \\
\tilde \Gamma^1_{21} &=& {1\over \sqrt{2} }(1- \eta^{AR})^{-1}
,\end{eqnarray}
so that $\tilde \Gamma^1_{21}$ is simply a diffusion operator,
as it is also manifest in Fig.\ref{fig9}. The explicit space and time dependence
is determined as
\begin{equation}\tilde \Gamma^1_{21}({ x};{ x}_1 , { x}_2 )=
{1\over \sqrt{2} }
 D_{t t_1}^{\eta=0}({\bf x}, {\bf x}_1)\delta({\bf x}_1 - {\bf x}_2) \delta(t_1-t_2)
.\end{equation}
The vertex $\Gamma^1_{12}$ has a more complicated structure due to the presence
of a Keldysh Green function in the kernel of the corresponding integral equation.
Its   detailed diagrammatic representation is shown in Fig.\ref{fig11}.
The integral $\eta^{RK}$ is given by
\begin{eqnarray}
\eta^{RK} & = &  
{1\over 2 \pi N_0 \tau} 
\int {\D^3 p \over ( 2 \pi )^3 }
G^R({\bf p} + {\bf q}/2 ) \nonumber\\
&&\label{eq42}
\times
G^K({\bf p}  +{\bf q}/2 -{\bf Q} ; {\bf p} +{\bf q}_1 -{\bf q}/2 -{\bf Q}).
\end{eqnarray}
We replace $G^K$ by $G^R (-\I/\tau) F G^A$, with
\begin{equation} \label{eq43}
F_{tt'}({\bf x}) = {1\over 2} \int {\D \hat p \over 4 \pi} 
g^K_{tt'}( {\bf p},{\bf x} )
\end{equation} 
and integrate over {\bf p} in Eq.(\ref{eq42}) to get
$\eta^{RK}_{tt'}({\bf x})  =  F_{tt'}({\bf x})$. 
It can be easily shown that $\eta^{KA}$ does not
contribute to the current; the (relevant part of the) vertex is then
found as
\begin{eqnarray}
\Gamma^1_{12}(x; x_1, x_2 ) &= & 
{1\over \sqrt{2}} 
\int \D \eta  D^{\eta}_{t_1-\eta/2,t-\eta/2}({\bf x}_1, {\bf x})\cr
&\times & F_{t, t-\eta}({\bf x}) 
\delta ({\bf x}_1 -  {\bf x}_2)
\delta (t_2 -t_1 +\eta )
.\end{eqnarray}
The last ingredient entering the expression of the interacting self-energy
is the electron-electron interaction propagator $V^{R,A}$.
At the level of the approximation we are working, it is sufficient
to confine to the  standard random phase approximation (RPA). 
The   retarded RPA screened Coulomb interaction reads :
\begin{eqnarray}
V^R(x,x') &= &V^0(x,x') \nonumber\\
&&\label{eq42a}
- \int \D x_1 \D x_2 V^0(x, x_1) \Pi^R(x_1, x_2 )V^R(x_2, x')
,\end{eqnarray}
where 
\begin{equation} 
V^0(x,x') = \delta(t-t')e^2/|{\bf x}- {\bf x'} |
\end{equation}
and
$\Pi^R (x_1, x_2)$ is the retarded component of the
density correlation function.
We write the density correlation function
as the sum of a ``static'' and a ``dynamic'' part, $\Pi^R = \Pi^s + \Pi^d$.
The ``static'' part
is given by
\begin{equation}
\Pi^{s} (x_1, x_2) = 2 N_0 \delta({\bf x}_1 -{\bf x}_2 )
\delta(t_1 - t_2 )
,\end{equation}
as it is seen directly from the quasi-classical expression for the electron density 
in Eq.(\ref{eq21}).
\begin{figure}
\begin{minipage}[t]{0.98\linewidth}
\hspace{0.5cm}{\epsfxsize=8.0cm\epsfbox{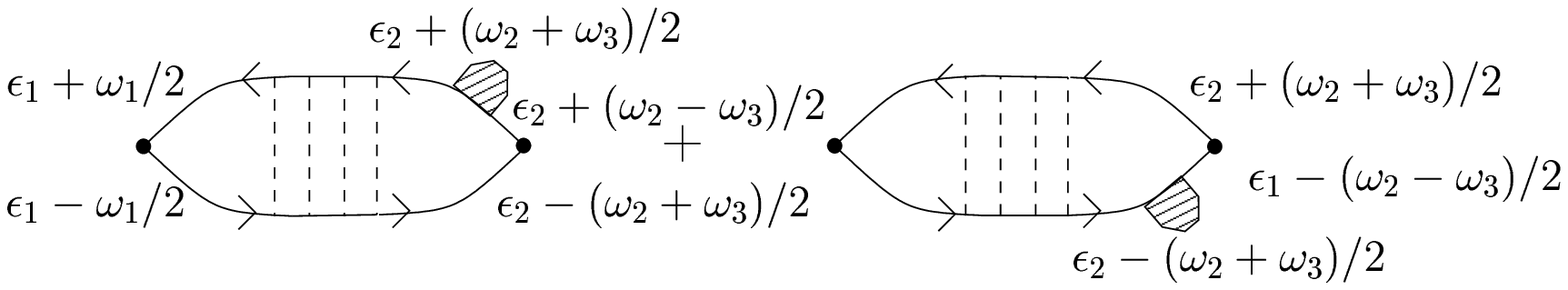}} 
\vspace{1cm}
\caption{The dynamical part of the
density correlation function.
}
\label{fig12}
\end{minipage}
\end{figure}
The ``dynamic'' part is determined 
from the ladder diagrams as
shown in Fig.\ref{fig12}. The expression is given by
\begin{eqnarray}
\lefteqn{ \Pi^{d}_{tt'}({\bf x}, {\bf x}' ) 
= 2\pi \I N_0 \tau
\int 
{\D\epsilon_1 \over 2\pi }
\cdots
{\D\omega_2 \over 2\pi }
\E^{- \I \omega_1 t + \I \omega_2 t' } } \nonumber\\
&& \times
D\left(  \epsilon_1+{\omega_1 \over 2}, \epsilon_1 - {\omega_1 \over 2};
    \epsilon_3 +{ \omega_2 \over 2}, 
    \epsilon_2 -{ \omega_2 \over 2} \right) \nonumber\\
&& \times \left[ 
F\left( {\epsilon_3+{\omega_2 \over 2} , \epsilon_2 +{\omega_2 \over 2}} \right)
- F\left({\epsilon_3-{\omega_2\over 2}, \epsilon_2 -{\omega_2 \over 2}} \right) 
    \right]
,\end{eqnarray}
where we suppressed the spatial indices. $D( \dots )$ is the diffuson and
$F$ is the angular average over $g^K$ as defined in (\ref{eq43}).
For the frequencies in $F$ we introduce the center-of-mass and relative
coordinates,
$F(\epsilon_1  , \epsilon_2  )
\to F_{(\epsilon_1+\epsilon_2)/ 2}( \epsilon_1-\epsilon_2 )$.
In order to evaluate the frequency integrals we assume that the
time dependent distribution function $F_\epsilon( t)$ deviates from 
the equilibrium distribution function only at small energies $\epsilon$,
i.e. $F_{\epsilon}(t) \to \pm 1 $ for
$\epsilon \to \pm \infty$.
We then find
\begin{equation} \label{eq46}
\Pi_{t_1 t_2}^{d}({\bf x}_1, {\bf x}_2) = 
 2 N_0 \tau{\partial \over \partial t_2} D^0_{t_1 t_2}({\bf x}_1, {\bf x}_2 ).
\end{equation}
Since here the diffuson enters with the relative time $\eta = 0$, 
the density correlation function does not depend on the external fields. 
Adding the static and dynamic parts together and
making use of the differential equation for the diffuson, one finds
\begin{equation}
\Pi^R( x_1, x_2 ) = - 2 N_0 \tau D  \partial_{{\bf x }_1}^2 
D^0_{t_1 t_2 }( {\bf x}_1, {\bf x}_2 ). 
\end{equation}
In the case of a uniform system  this reduces to the
standard expression
\begin{equation}
\Pi^R( {\bf q}, \omega  )  
= 2 N_0 {Dq^2 \over -\I \omega + Dq^2 }
\end{equation}
from which the dynamically screened Coulomb interaction is
determined as
\begin{eqnarray}
V({\bf q}, \omega ) &=&{ V^0(q) \over 1 + V^0(q) 2N_0 Dq^2/(-\I \omega + Dq^2) }\\
\label{eq59}
&\approx & {1\over 2 N_0}  {-\I \omega + Dq^2 \over Dq^2 }. 
\end{eqnarray}
The second line is valid, when $|\omega | < V^0(q) 2  N_0 Dq^2$.
In three dimensions, where $V^0(q) = 4\pi e^2/q^2$, this condition
reads $|\omega| < D \kappa^2 $. Since the inverse screening length
$\kappa$ is in a metal typically of the order of the Fermi wavelength
the approximation is well justified. 
In lower dimensions a more careful analysis is sometimes necessary,
see for example Ref.\cite{blanter96,raimondi99}.

If we now collect all the pieces of our analysis we may 
come back to the quantum correction to
the current. For convenience we 
switch from the energy/momentum domain to the time/space domain.
Remember that 
we have neglected the energy dependence of the Green function $G^R$ and $G^A$
in the calculation of the Hikami box.
This is equivalent to approximate these Green functions as
local in time.
The resulting time dependencies are shown in Figs.\ref{fig10} and 
\ref{fig11}.
The correction to the current 
$\delta {\bf  j} = \delta {\bf j}_a + \delta {\bf j}_b $
is finally found as
\begin{eqnarray}
\label{currentformula}
\delta {\bf j}_a ({\bf x}, t) &=& - D \nabla \delta \rho({\bf x},t)\\
\delta {\bf j}_b({\bf x}, t ) &=&
e 2\pi  D N_0  \tau^2
\int \D \eta \D x_1 \D x_2 \nonumber\\
&&\times
F_{t-\eta ,t}({\bf x}) D^\eta_{t-\eta/2, t_1 -\eta/2}({\bf x}, {\bf x}_1  ) 
F_{t_1, t_1-\eta }({\bf x}_1)
\nonumber\\
&& \times V^R_{t_1, t_2}({\bf x}_1, {\bf x}_2)
(-\I \nabla_{\bf x}) D^0_{t_2, t-\eta}({\bf x}_2, {\bf x} ) 
+ c.c. .
\label{eq61}\end{eqnarray}
This current formula is one of the central results of our paper.
The nice feature is that is valid for arbitrary form of the distribution function
and diffuson propagator. This will allow us to examine the current in different
experimental and geometrical setups as well as more general questions concerning
its physical interpretations. Specific 
applications are discussed in Sections \ref{sectionGauge},\ref{sectionWire}. 
The interacting disordered electron problem is often formulated in terms
of the field-theoretic non-linear sigma model and one may wonders how the
presented diagrammatic approach is related to it.
To this end we explicitly show in the appendix \ref{sectionSigma} at the end of the paper that
the same expression for the current can be obtained  
from the field theoretic approach of Ref.\cite{kamenev99}.
%
%
%
\section{Gauge invariance}
\label{sectionGauge}
As a first application of the formula for the current we discuss in this
section the issue of its physical interpretation and  of gauge invariance.
To begin with, we believe useful to make contact
with our earlier work in Refs.\cite{raimondi99,leadbeater00},

In Refs.\cite{raimondi99,leadbeater00}
the quantum correction to the current was derived working in a vector gauge,
${\bf A}=-t {\bf E}$, $\phi =0$,  under the assumption that the
electron distribution function had the equilibrium form, $F(\epsilon, {\bf x} )= \tanh(\epsilon/2T)$,
or, equivalently, in the time domain 
$F_{tt'}({\bf x}) = - \I T /\sinh[ \pi T (t-t') ] $, and that
we dealt with  a uniform system with a homogenous 
charge density. In this situation one has immediately that  
$\delta {\bf j}_a = 0$ and by Fourier transforming Eq.(\ref{eq61}) with respect to
the spatial variables,  the current formula of \cite{raimondi99,leadbeater00} is reproduced.

In \cite{raimondi99,leadbeater00} we showed that in the limit of weak electric field
our theory reproduces the well-known Altshuler-Aronov corrections to the conductivity.
At larger fields  non-linear contributions to the current arise as a consequence of the  
nonlocal character of the current formula. 

By working in the vector gauge, the  electric field enters the equation for the diffuson,
Eq.(\ref{diffusonequation}), via the minimal substitution of the vector potential.
Due to the quasi-classical nature of the equation governing the diffuson, one may
interpret the non-linear conductivity  in terms of phases. Let us recall the argument
of Ref\cite{leadbeater00}.
The interaction correction to the conductivity is related to the propagation of a
particle and a hole along closed paths.
Pictorially one may think of this as one particle going around
a closed path, starting for example at $t=0$ and arriving at
$t=\eta$. This particle is also interacting with a background particle
which is retracing backwards-in-time
the same closed path. Since the point of interaction ${\bf x}({t_1})$ can be
anywhere along the path, the particles traverse the loop at
different times.
In the presence of a vector potential the accumulated phase difference of the two paths
is
$\varphi_1 - \varphi_2 =  
  e \int_{t_1-\eta}^{t_1} \D t' \dot{\bf x}_1 \cdot {\bf A } 
  - e \int_{0}^\eta  \D  t'     \dot{\bf x}_2 \cdot {\bf A }$.
This can be simplified using that ${\bf x}_1(t)={\bf x}_2(t)$ for
$0<t < t_1$ and ${\bf x}_1(t-\eta)={\bf x}_2(t)$ for $t_1 <t<\eta$
leading to
$\varphi_1 - \varphi_2 =
e \int_{t_1-\eta}^0 d t' \dot{\bf x}_1 \cdot[ {\bf A}(t')- {\bf A}(t'+\eta) ] $.
For the particular case of a static electric field described by ${\bf A} = - {\bf E} t$, the
above given phase difference becomes
$\varphi_1 - \varphi_2 = e \eta ({\bf x}_2 - {\bf x}_1 ) \cdot {\bf E}$.
This suggests that the interaction correction should be sensitive to a static
electric field, leading  to a non-linear conductivity.

One may object against this interpretation by observing 
 that the vector potential can be gauged away in 
such a way that the static electric field is described by   a static 
scalar potential $E({\bf r}) = - \nabla \phi({\bf r})$.
A static scalar potential, again according to Eq.(\ref{diffusonequation}),
does no longer affect the diffuson propagator, so that    
the argument of the phase difference  along the two paths cannot be used.
Of course this does not imply that the current formula is incorrect.
 
In fact we may demonstrate explicitly  the gauge invariance of the current formula.
First one notices that  $\delta {\bf j} _a = -D \nabla \rho$ is gauge invariant.
For $\delta {\bf j}_b$ an explicit check is necessary.
Given the gauge transformation
\begin{eqnarray}
{\bf A} & \to & {\bf A} +\nabla \chi \\
\phi &\to & \phi - \partial_t \chi 
\end{eqnarray}
the diffuson and the distribution function  transform according to
\begin{eqnarray}
F_{tt'}({\bf x} )  & \to & 
F_{tt'}({\bf x})
\exp\{ -\I e[ \chi({\bf x}, t)- \chi({\bf x}, t')]  \} \\
D_{tt'}^{\eta}({\bf x}, {\bf x}' ) & \to & 
D_{tt'}^{\eta}({\bf x}, {\bf x}' )	      
\\
&& \times \exp\{ -\I e [ \chi({\bf x}, t+{\eta\over 2} ) - \chi({\bf x}, t-{\eta \over 2})] \}\nonumber \\
&& \times \exp\{ \I e [\chi({\bf x}', t'+{\eta\over 2})- \chi({\bf x}', t'-{\eta \over 2})]\} 
.\nonumber \end{eqnarray}
By applying the above transformation to Eq.(\ref{eq61}), one easily verifies
that the function $\chi({\bf x}, t)$ drops, 
so that  the expression is manifestly gauge invariant.
For the special example of a static electric field with ${\bf A} = - {\bf E} t$ 
we choose
$\chi({\bf x},t) = t  {\bf E} \cdot {\bf x}$. 
After the gauge transformation the electric field
appears in the distribution function
$\tanh(\epsilon/2T) \to \tanh[(\epsilon - \mu_{\bf x} )/2T]$, 
$\mu_{\bf x} =e {\bf E} \cdot {\bf x}$ but not in the diffuson.

We conclude that although the correction to the current as derived in this
paper is gauge invariant, the interpretation of the non-linear effects depends
on the actual choice of the potentials ${\bf A}$ or $\phi$.
With ${\bf E} = - \partial_t {\bf A}$, $\phi= 0$ we would interpret the non-linear conductivity 
as due to dephasing. With ${\bf E} = - \nabla \phi$, ${\bf A}=0$ the reason of the non-linear
conductivity is attributed to  the different local chemical potential 
felt by  the particle and hole.

\section{Mesoscopic wire}
\label{sectionWire}
In this section we use the current formula to analyze 
the non-linear electrical transport
in a thin wire.
We assume that the temperature is low enough so that the Drude conductivity
is dominated by the impurity scattering and is therefore temperature independent.
It is well known 
that in one dimension the electron-electron interaction  leads to a $1/\sqrt{T}$
correction to the conductivity. The interesting question to ask concerns  
what happens at larger voltages  and which are the relevant length and energy 
scales in the problem.

For the calculation we need the diffuson propagator and the distribution function in the wire.
At the boundary with the vacuum or an insulator the derivative of the diffuson normal
to the boundary vanishes,
$( {\bf n}\cdot \nabla ) D({\bf x},{\bf x}') =0$.
In the case of an infinitely long wire with cross section $S$
the solution of the diffusion equation reads
\begin{equation}
D( x, t) = {1\over \tau}{1\over S}{1\over \sqrt{ 4 \pi D t} } \exp [-  x^2/(4 Dt  )].
\end{equation}
In the above result we have averaged the diffuson over the cross
section. 
In a wire of finite length we impose the open boundary conditions along the
$x$ axis
\begin{equation}
D_{tt'}( x,  x') \big{ |}_{  x,  x' = 0,L } =0 
,\end{equation}
corresponding to the fact that an electron arriving  at the boundary escapes
into the leads, where dissipation takes place. Therefore it no longer contributes to the
phase coherent process of quantum  interference.
The diffuson in the finite system is related to the  propagator for an infinite system according to
\begin{eqnarray}
D_{tt'}( x,  x')& = & \sum_{n= - \infty}^{\infty}\big[
D( x - x' + 2 n L, t-t' ) \nonumber\\
\label{eq68}
&& - D( x +  x' + 2 n L, t-t')  \big].
\end{eqnarray}
For the actual calculations it is convenient to consider also the product of the
retarded interaction and the diffuson,
\begin{equation} \label{eqVRD}
(V^R D)_{tt'}({\bf x}, {\bf x}')
 = \int \D {\bf x}_1 \D t_1 V^R_{t t_1}({\bf x}, {\bf x}_1 ) D^0_{t_1 t'}({\bf x}_1, {\bf x}')
.\end{equation}
For the case of long range interaction this product solves the equation
\begin{equation}
[ -D_{} \nabla_{\bf x}^2 ] (V^R D)_{tt'}({\bf x}, {\bf x}') = {1\over 2 N_0 \tau } 
\delta({\bf x } -{\bf x}') \delta(t-t') 
,\end{equation}
as it may be seen by comparing with Eq.(\ref{eq59}).
For a one dimensional wire with open boundary conditions we obtain then
\begin{eqnarray}
( V^R D)_{tt'}( x , x')& =& 
\left[ {(L-  x')  x \over L } - ( x- x')
 \Theta( x- x') \right] \nonumber\\
&&\times {1\over 2 D_{} N_0 \tau} \delta(t-t')
.\end{eqnarray}
Besides the boundary conditions for the diffuson and the interaction we need the
boundary condition for the distribution function.
We  assume that the leads of the wire are in thermal equilibrium, so that
the distribution function is given by 
\begin{equation}F(\epsilon ,{\bf x})\big{|}_{{\bf x}= 0,L }= 
\tanh\left( {\epsilon \pm eV /2 \over 2 T} \right)
.\end{equation}
The distribution function inside a mesoscopic wire has been investigated
both experimentally\cite{pothier97} and theoretically\cite{kozub95,naveh98}.
In the theoretical analysis, in particular,  a solution of  the Boltzmann equation in the presence of disorder, 
electron-electron
interaction and electron-phonon interaction has been given.
Here we borrow the approximate solutions for the
the distribution function found in Refs.\cite{kozub95,naveh98}.

The form of the distribution function depends on the various relaxation mechanisms
governing the collision integral.
In the following we  first consider a long wire, $L \gg L_{\rm ph}$, and
then subsequently reduce the length to $L_{\rm ph} \gg L \gg L_{\rm in}$ and
$L_{\rm in} \gg L \gg L_T$. Here we indicate with
$L_{\rm ph}$, $L_{\rm in},$ and $L_T$  the 
electron-phonon, the inelastic and the thermal scattering lengths.

\subsection{Long wire}
In the case of  a long wire, $L \gg L_{\rm ph} \gg L_{\rm in}$,
the electrons which traverse the wire scatter many times inelastically and exchange energy with the
environment, for example with the phonons.
As a result the distribution function  acquires  the equilibrium form 
with a local chemical potential
and temperature. Our ansatz for the distribution function is
\begin{equation} \label{eq73}
F({\epsilon, x }) =
\tanh\left( {\epsilon +e V(L-2x)/2L \over 2 T_{e}(x) } \right)
.\end{equation}
We assume that the temperature is constant in the bulk of the wire, 
and we also neglect  the
region near the leads where the electron temperature increases from 
the value in the leads to the one in the bulk.
The electron temperature in the bulk may be estimated with standard energy balance arguments
\cite{bergmann90,altshuler98}.
For  a stationary temperature $T_{e}$,
the Joule heating power $P_{\rm in}= \sigma {\bf E}^2 $ equals the power which 
is transferred into the phonon
system, $P_{\rm out}$.
For weak heating, one has 
$P_{\rm out}= c_V \Delta T/ \tau_{\rm ph}$,
where $c_V$ is the electron specific heat.
One obtains then that the difference of electron and phonon temperature may be estimated as
\begin{equation} \label{eq145}
\Delta T \approx {3\over \pi^2 } D (e V/L)^2 \tau_{\rm ph} /T.
\end{equation}
For strong heating, on the other hand, the effective electron temperature is of the order of the voltage drop 
over a phonon length
\begin{equation}
T_e \sim e V L_{\rm ph}/L.
\end{equation}

We are now ready to evaluate the  quantum correction to the current in the wire as
\begin{equation}
I ={1\over L} S \int_0^L \D x j^x( x )
,\end{equation}
where $j^x$ is the component of the current parallel to the wire.
Recall that we separated the correction to the current density into two contributions
$\delta {\bf j} = \delta {\bf j}_a + \delta {\bf j}_b$.
The first of the two terms does not contribute to the correction to the
current since
\begin{eqnarray}
\int_0^L \D x \delta { j}_a & \propto & \int_0^L \D x {\partial  \over \partial x  } 
\delta g^K(x) \\
 &=& \delta g^K( L) - \delta g^K(0)
\end{eqnarray}
and the boundary conditions impose that $\delta g^K$ vanishes on the leads.
The correction to the current  then reads
\begin{eqnarray}
I &=&{ 2 \pi e \tau  \over L}  \int_0^L \D x \D x_1 \int \D \eta 
\Re \left\{  
 \vphantom{ x \over L} 
F_{t-\eta,  t}(x) 
F_{t-\eta, t-2 \eta}(x_1) \right. \nonumber \\
&& 
\label{eq150}
\times 
\left.
D^\eta_{t-\eta/2, t-3\eta/2}(x, x_1) 
(-\I) \left[ \Theta(x_1 - x) - { x_1 \over L} \right] \right\}
.\end{eqnarray}
It is useful to introduce the center-of-mass and relative coordinate $R=(x+x_1)/2$, $r=x_1-x$.
The last term in Eq.(\ref{eq150}) above becomes then
$\Theta(r) - R/L - r/2L $. The term $R/L$ vanishes upon integration due to the
antisymmetry of the $r$-integral. 
The assumption that the thermal length is much shorter than the system size
allows furthermore to approximate $
\Theta(r) - r/2L $ by $\Theta (r)$,
to neglect the boundary effect on the diffuson, i.e $
D^\eta_{t-\eta/2, t-3\eta/2}(x, x_1) \approx D(r,\eta)$,
and to extend the $r$-integration to infinity.
By inserting the distribution function we arrive at
\begin{eqnarray}
\delta I &=& - 2\pi e \tau  \int_0^\infty \D r \int_0^\infty \D \eta  
\left({ T_{ e} \over \sinh(\pi T_{ e} \eta ) } \right)^2 
\nonumber\\
&&
\label{eq151}
\times D(r,\eta) 
\sin( {e V r \eta /L} )
.\end{eqnarray}
This is equivalent to what is obtained 
in Ref.\cite{raimondi99}.
In the limit of low voltage the current becomes
\begin{equation}
\delta I(V) \approx 
{e^2 \over \pi^2 } {\sqrt{ D/ T_{e}} \over L} V
\left(-4.92 + 0.21 { D( e V/L)^2 \over T_{ e}^3 } + \cdots \right)
,\end{equation}
where $4.92$ and $0.21$ are approximate numerical factors.
In order to obtain the full current, one has to add the contribution of the
Drude leading term, i.e., $I = 2 e^2 D N_0 S V/L + \delta I$. 
The low voltage expansion applies when the voltage drop over a thermal length is smaller
than the temperature, $ e V L_{T}/L < T_e $. Since we have assumed that 
$L_{\rm ph} \gg L_T $, the electron temperature as a function of voltage rises so fast that
the condition always holds.
Finally we compare the heating and non-heating contribution to the non-linear conductivity at low voltage.
By taking the linear conductivity 
and the increase in temperature due to low voltages from Eq.(\ref{eq145}),
we find the heating contribution to the
cubic term in the current voltage characteristics in the form
\begin{equation}
\delta I_{\rm heating} \approx
4.92 { e^2 \over \pi^2} {L_T \over L }  V{3\over 2 \pi^2} 
{ D(e V/L)^2 \over T^2 (1/\tau_{\rm ph}) }
.\end{equation}
This has to be compared with the corresponding non-heating cubic contribution
\begin{equation}
\delta I_{\rm non-heating} \approx 0.21 {e^2\over \pi^2} {L_T \over L} V  { D (eV/L)^2 \over  T^3}
.\end{equation} 
One observes that the heating contribution is by a factor of the order of $ T\tau_{\rm ph}$
larger than the non-heating contribution.

\subsection{Intermediate length}
\begin{figure}
\noindent
\begin{minipage}[t]{0.98\linewidth}
\hspace{0.5cm}{\epsfysize=4.0cm\epsfbox{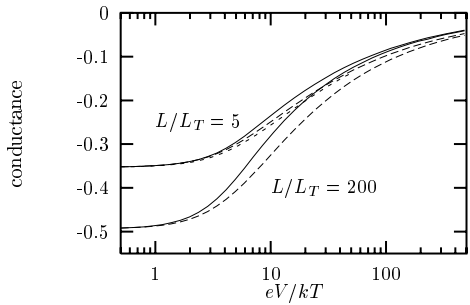}} 
\caption{Interaction correction to the conductance $I/V$ for a mesoscopic wire 
as a function of voltage. $I/V$ is plotted in units
of $(e^2/\hbar) L_T/L$.  
The full line corresponds to the non-equilibrium distribution function (\ref{eqNonequi}). 
The long dashed line corresponds to
the local equilibrium distribution function (\ref{eq73}) with the $x$-dependent temperature.
The short dashed line ($L/L_T = 5$) 
is the non-linear conductivity due to the heating contribution only, Eq.(\ref{eqHeating}).}
\label{figIV}
\end{minipage}
\end{figure}

Now we consider a wire of  intermediate length where $L_{\rm ph}\gg L \gg L_{\rm in}$.
One still expects to be near local equilibrium
although  the main mechanism which carries the energy out of the wire is not due to the phonons, 
but  to the heat flow out of the wire.
Under these conditions, a temperature profile over the wire develops. The local temperature 
satisfies the equation\cite{nagaev95}
\begin{equation}
{ \D^2 \over \D x^2} T^2_e(x)  = - {6 \over \pi^2} { ( e V )^2 \over L^2 }
\end{equation}
which is solved by
\begin{equation} \label{eqTx}
T_{ e}^2(x) = T^2 + { 3 \over \pi^2 } \left( {e V \over L } \right)^2 
x (L-x ).
\end{equation}
The correction to the current at a point $x$ probes the wire in a region $\sim L_T$ around $x$.
As a consequence, non-linearities in the $I-V$ characteristics  arise because
(1) the temperature depends on the voltage (``heating''), and of
(2) the non-local character of the current formula (``non-heating'').
It turns out that the first effect dominates, whereas the non-heating contribution to the 
non-linear conductivity is only a small  perturbation.
For illustration,  notice that the heating becomes strong at $e V \sim kT$.
Non-heating non-linearities, on the other hand, arise on the scale $e V L_T/L \sim k T$. 
Since we assume that $L \gg L_T$,  heating is indeed dominant.
This is also demonstrated in Fig.\ref{figIV} where we plot the conductance
$ \delta G = \delta I/V$ as a function of voltage,
while varying the system size, $L/L_T$, and the distribution function
$F_{tt'}(x)$.
For large $L/L_T$ the linear conductance approaches 
$\delta I/V \approx - (4.92 e^2 /\pi^2)(L_T/L) V $.
For the smaller system size $L/L_T =5$, the linear conductance is
suppressed, due to the chosen boundary conditions.
The long dashed lines and the full lines correspond to the local equilibrium distribution (\ref{eq73}), and
 to the non-equilibrium distribution discussed below, respectively.
The short dashed line ($L/L_T=5$), instead,  is obtained by taking into account only the heating contribution,
 i.e. by  calculating from Eq.(\ref{eq150}) the {\em linear} conductivity and averaging over 
the $x$-dependent temperature,
\begin{equation}\label{eqHeating}
\delta I_{\rm heating} = {1\over L} \int_0^L \D x \delta \sigma ( T(x) ) (V/L)
.\end{equation}
One observes in Fig.\ref{figIV} that the ``heating'' contribution
reproduces with a  good accuracy the much more complicated full calculation.
For the longer system with $L/L_T=200$,  we do not plot the ``heating'' curve,
because in this case it is practically indistinguishable from the  long dashed one.
A slightly larger non-linear conductivity is found in the non-equilibrium situation 
as it will be discussed in the next section.

\subsection{Short wire}
In a very short wire the inelastic length may exceed the system size $L$.
When one neglects the  inelastic scattering,  the distribution function inside the wire becomes 
a linear superposition of the distribution functions in the leads and reads
\begin{equation} \label{eqNonequi}
F(\epsilon, x) =  \left[ (L-x) F(\epsilon, 0 ) + x F(\epsilon, L)\right]/L
.\end{equation}
In the limit $L \gg L_T$ the analytic calculation of the current proceeds as in the case 
of subsection A. In analogy
to (\ref{eq151}) we arrive at
\begin{eqnarray}
\delta I  &= &- 2 \pi e \tau \int_0^\infty  \D r \int_0^\infty \D \eta
\left({ T \over \sinh( \pi T \eta ) } \right)^2 \cr
&& \times D(r, \eta) \sin( e V \eta ) r/L
.\end{eqnarray}
The numerical results for the
current-voltage characteristics in the presence of such a distribution function are
shown in Fig.\ref{figIV}. Notice that we integrated numerically Eq.({\ref{eq150}),
as it is appropriate when $L_T/L$ is not very large.
The linear conductance is -- of course -- the same as in the local equilibrium situation, 
whereas the non-linear effects are slightly larger.

\subsection{The Spin-triplet channel}
Up to now we have neglected all the spin effects.
As demonstrated in appendix \ref{sectionTriplet},  
the current formula can easily be generalized in order to 
include also the so called spin triplet channels.
The general equation for the correction to the current is given in the appendix.
With the two-step distribution function (\ref{eqNonequi}) and for
$L \gg L_T$ 
the equation for the current reads  
\begin{eqnarray}
\delta {\bf j}_b &= &- 4 \pi e D_{}N_0 \tau^2 \sum_{i=0}^3 
\int_{-\infty}^{\infty} \D r \int_0^{\infty} \D \eta \int_0^\eta \D t_1 \cr
&\times & \left({T \over \sinh( \pi T \eta)} \right)^2 \sin ( e V \eta ) {r\over L} \cr
&\times & D(r, t_1) { \partial \over \partial r} ( \tilde \gamma^i D )(r, \eta -t_1 )
\label{eqCurrentTriplet}
,\end{eqnarray}
with
\begin{equation}
(\tilde \gamma^i D)(r,t) = { \gamma^i \over \tau } \sqrt{{1- 2\gamma_i \over 4 \pi D t}}
\exp\left[- r^2 (1-2 \gamma^i )/4Dt \right]
,\end{equation}
and $\gamma^i$ are the interaction amplitudes in the spin singlet ($\gamma^0=1/2$) 
and spin triplet ($\gamma^{1,2,3}= \gamma^t$) channels.
In the low voltage limit this expression reproduces the standard result for the
linear conductivity,
\begin{equation}
\delta \sigma = -4.92 {e^2 \over \pi^2}{L_T \over L} 
\left[1 - 3(\sqrt{1-2\gamma_t} -1 +\gamma_t)/\gamma_t  \right]
.\end{equation}
Fig.\ref{figIV_triplet} depicts the current voltage characteristics 
for different strengths of the triplet scattering amplitude.
Again we compare the full result (\ref{eqCurrentTriplet}) 
with the simple heating contribution, that is the average of the linear conductivity
over the temperature profile given in (\ref{eqTx}).
\begin{figure}
\noindent
\begin{minipage}[t]{0.98\linewidth}
\hspace{0.5cm}{\epsfxsize=5.0cm\epsfysize=4.0cm\epsfbox{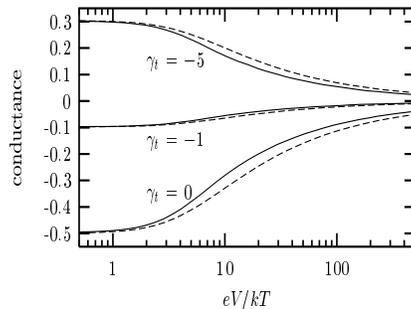}} 
\caption{Correction to the conductance as a function of voltage.
Notice that for a strong negative scattering amplitude in 
the spin triplet channel, $\gamma_t$, the quantum correction changes sign. The dashed line corresponds
to the hot electron regime (heating), while the full line  to the non-equilibrium situation.}
\label{figIV_triplet}
\end{minipage}
\end{figure}
For strong scattering in the triplet channel the quantum correction changes sign. 
As a function of voltage, the quantum corrections are suppressed.
For the case of a distribution function out of equilibrium,
 the non-heating non-linear contributions  are stronger than the pure heating effects.
\section{Conclusions}
We have calculated the interaction correction to the electrical current for a disordered metal out of equilibrium. 
In order to do so we have extended the diagrammatic approach of Refs.\cite{raimondi99,leadbeater00}.
We have first demonstrated explicitly the gauge invariance of our current formula
for the current density. In particular, we have discussed how the physical interpretation
of the non-linear contribution may be described differently depending on the gauge choice made.
 In the scalar gauge 
one may think as the particle-hole pair lying on a different chemical potential. In the vector gauge, the 
chemical potential of the particle and hole are equal. In the latter case an argument based on phase differences
leads to the  conclusion that the electric field affects the correction to the conductivity.

We have successively discussed in some detail the correction for the case of  a mesoscopic wire.
We have assumed that the wire is attached to ``ideal leads'' (infinite conductance), 
which are described by means of  boundary
conditions for the distribution function and the diffusion propagator. 
We have distinguished three different regimes.
First we concentrated on a wire near local equilibrium, with a constant electron temperature.
Besides heating, there is a contribution to the non-linear conductivity due to the nonlocal nature of the
current response.

For short wires, where the electron-phonon length is longer than the system size, we found that the
conductance scales with voltage over temperature. 
Such a scaling behavior has been recently observed in a nanobridge\cite{weber00}.
However, the quantitative shape of the ``scaling curve'' is not universal. We have found  two different
curves in the hot electron regime, where the inelastic scattering length is shorter than the system size,
and in non-equilibrium.

We would like finally to emphasize that in this
paper we have concentrated on the quantum correction to the current, and we have left aside
a derivation of  
the charge density
and the kinetic equation 
including the quantum corrections. This latter task implies the evaluation
of the quantum correction to the distribution function. 
The evaluation of the distribution function and charge density 
are directly connected  as it is clear 
when one expresses the
distribution function in terms of the quasi-classical Green functions as
$g^K = g^RF -F g^A $, and recalls the relation of $g^K$ with the charge density.
The inclusion of quantum corrections into the
kinetic equation for disordered electrons have been considered
for the weak localization case without electron interaction in
Refs.\cite{hershfield86,strinati89},
and in the presence of interactions in Ref.\cite{strinati91}.
In the latter work, the non-linear electric field effects have not been included.
We notice that the quantum correction to the charge density out of equilibrium can be derived following
the procedure described in this paper for the current density in section III. 
This task is  however   more involved for the following reason.
Within the leading order in the gradient expansion, one finds from 
$\delta \Sigma^{{\rm imp}, K}$, in analogy to Eq.(\ref{eqj_a}), 
the identity $\delta \rho_a  = \delta \rho$.
The diagrams in Fig.\ref{fig10}, responsible for the contribution to the current denominated
$\delta j_b$, give in the case of the charge density
$\delta \rho_b=0$. This happens because,  the evaluation of  the Hikami box with the density vertex is zero 
in the leading order of the gradient expansion. In order to consider the next-to-leading terms,
one has to take into account higher powers in  the inverse of $\epsilon_F \tau$ and $ql$.
This requires the evaluation of also the diagrams with only one
vertex, $\Gamma$ or $\tilde \Gamma$,  renormalized by the diffusion pole. Therefore
the full expression of the electron self-energy is considerably more complicated and  
such a calculation will be more lengthy than the one we presented here for the current density.
This task, although worth to be done,  is beyond the scope of the present paper.

\acknowledgements
We acknowledge many discussions with C. Castellani. 
This work was supported by the
DFG through SFB 484 and Forschergruppe HO 955.
R.R. acknowledges partial financial support from
E.U. under Grant number RTN 1-1999-00406.
\begin{appendix}
\section{Field theoretic approach: the non-linear sigma model}
\label{sectionSigma}
The field theoretic formulation of the interacting, disordered electron
system was pioneered by Finkelstein in the 80's\cite{finkelstein83}.
Recently,  this formulation has been extended to the non-equilibrium case by means of  the Keldysh technique
by Kamenev and Andreev\cite{kamenev99} and Chamon, Ludwig and Nayak
\cite{chamon99} for the case of normal-conducting metals and by 
Feigel'man, Larkin and Skvortsov \cite{feigelman99} for superconductors.
Gutmann and Gefen \cite{gutmann00} have also used the field theoretic description
to  calculate the current and zero frequency shot noise.

Given the already extensive  literature available on the subject, we 
believe that, rather than repeating again  the derivation of the 
non-linear sigma model,  it is perhaps more useful,
instead,  to show  how to obtain our current formula (\ref{eq61}) within the
non-linear sigma model. In this appendix we will discuss the spinless version of the
model, following Ref.\cite{kamenev99}, postponing the spin effects to the following appendix.
In the absence of the electron interaction, the action of the non-linear sigma model is
given by
\begin{equation}
\I S_0 = -{\pi N_0 \over 4}
\left[ D \Tr (\partial_{\bf x} Q)^2 + 4\I  
\Tr \epsilon Q  \right],
\end{equation}
where the so-called long derivative $\partial_{\bf x}$ is defined by
\begin{equation}
\partial_{\bf x} Q = \nabla Q + \I e [ {\bf A}, Q ]
.\end{equation}
The field $Q\equiv Q^{ij}_{tt'}({\bf x})$ must satisfy the  constraints
 $Q^2=1$ and $\Tr Q =0$.
The electron interaction is described by  the following term in   the action 
\begin{equation}
\I S_1 = -\I \pi N_0 \Tr \Phi_\alpha \gamma^\alpha Q 
,\end{equation}
with $\gamma^1 = \sigma_0$ and $\gamma^2 = \sigma_x$.  
The fluctuations of the field $\Phi$ are
related to the statically screened Coulomb interaction and given by
\begin{equation}
-\I \langle \Phi_i({\bf x},t ) \Phi_j({\bf x}',t') \rangle = {1\over 2 } V \sigma_x^{ij} 
 \delta( {\bf x} - {\bf x}') \delta(t-t').
\end{equation}
In the case of long range Coulomb forces one finds
$V = 1/N_0 $.
Since $\Phi$ couples only linearly to $Q$ 
it can be integrated out:
\begin{eqnarray}
\lefteqn{ \langle \E^{- \I \pi N_0 \Tr \Phi_\alpha \gamma^\alpha  Q } 
\rangle_{\Phi} }&& \nonumber\\
\label{eqA5}
&&= \exp\left[ -{\I  V (\pi N_0)^2 \over 2}
\int \D{\bf x} \D t \Tr (\gamma^1 Q_{tt}({\bf x} )) 
\Tr ( \gamma^2 Q_{tt}( {\bf x} ) )
\right].
\end{eqnarray}
Here the trace refers to the Keldysh space.
The appearance of  the product of terms containing $\gamma^1$ and $\gamma^2$ 
stems from the $\sigma_x$ structure  of the interaction
matrix in Keldysh space.
In order to make contact with the diagrammatic approach, we express the
Green functions in terms of the  $Q$-fields.
First we observe that \cite{kamenev99}
\begin{equation}
\hat G_{tt'}({\bf x}, {\bf x'} )=
\langle \left[G_0^{-1} +{ \I \over 2 \tau} Q +
\Phi_\alpha \gamma^\alpha  \right]^{-1}  \rangle_{Q,\Phi} 
,\end{equation}
where the brackets $\langle \dots \rangle_{Q,\Phi}$ 
indicate that one has to average over the fields $Q$ and $\Phi$.
For the sake of simplicity, we drop the subscripts $Q$, $\Phi$ in the following.
By using the condition that the fields $Q$ and $\Phi$ are only slowly
varying in space and time, one finds a relation between the $\xi$-integrated
Green function and the $Q$-matrix.
In particular, upon taking the  $s$-wave component of both, one finds
\begin{equation}
 \int {\D \hat p \over 4 \pi } \hat g_{tt'} ({\bf {\hat p}}, {\bf x} ) = 
 \langle \hat Q_{tt'}({\bf x}) \rangle
.\end{equation}
In a similar way the $p$-wave part of the $\xi$-integrated Green function
is related to $Q$ by
\begin{equation} \label{eq76}
v_F \int {\D \hat p \over 4\pi } {\bf {\hat p}} \hat g_{\epsilon \epsilon'} 
({\bf {\hat p} }, {\bf x} )
= {1 \over 2} D 
\langle \partial_{\bf x} Q Q - Q \partial_{\bf x} Q \rangle
,\end{equation}
so that  the current density reads
\begin{eqnarray} 
\label{eq78}
{\bf j} &=& {1\over 2} e \pi D N_0 
\langle (\partial_{\bf x} Q Q -Q\partial_{\bf x}Q)^{12}\rangle  
.\end{eqnarray}
At a first glance, this might differ from what is found 
following Ref.\cite{kamenev99}, 
where the current density is written as
\begin{eqnarray} \label{eq80}
{\bf j} &=& {1\over 2} e \pi D N_0 
\langle \Tr \gamma^2 (\partial_{\bf x} Q Q -Q\partial_{\bf x}Q)\rangle  
.\end{eqnarray}
On the other hand, by comparing with the expression (\ref{eq76}) for the Green function, 
it is seen that
Eq.(\ref{eq80}) sums the Keldysh component and the $(21)$-component of the Green function,
whereas (\ref{eq78}) takes only the Keldysh component. 
Since the $(21)$-component is zero the two expressions are equivalent.

\subsection{Propagators}
The saddle point approximation for the  $Q$-field 
\begin{equation}
Q^{\rm sp} = \left( \begin{array}{cc} 1 & 2F \\ 0  & -1
                \end{array} \right)
\end{equation}
reproduces the Drude-Boltzmann
theory. The quantum corrections are found when considering the
fluctuations about the saddle point.
We parameterize $Q$ according to
\begin{equation}
Q= u \E^{-W/2} \sigma_z \E^{W/2} u 
\end{equation} 
where $u$ characterizes the saddle-point distribution function
\begin{equation}
u= \left( \begin{array}{cc}
1 & F \\
0 & -1 
\end{array} \right), \,
Q^{\rm sp} = u\sigma_z u 
,\end{equation}
and $W$
parameterizes the fluctuations,
\begin{equation}
W = \left( \begin{array}{cc}
0 & w \\
\bar w & 0 \end{array} \right).
\end{equation}
By expanding  in powers of the $W$-field, one finds
for the non-interacting action, $S_0$, up to the quadratic order
\begin{eqnarray}
\I S_0^{(2)} &=& -{\pi N_0 \over 2 }
\left\{ \int \D {\bf x} \D t_1 \D t_2
w_{t_1 t_2}
\left(
 \partial_{t_1} + \partial_{t_2} + D \partial_{\bf x}^2 \right) 
 \bar w_{t_2 t_1}  \right. \nonumber \\
&+& \left.   \int    \D{\bf x} \D t_1 \cdots \D t_2' 
 \bar w_{t_1 t_2} (\partial_{\bf x} F_{t_2 t_2'} ) 
  \bar w_{t_2' t_1'} ( \partial_{\bf x}  F_{t_1' t_1} )\right\}
.\end{eqnarray}
The long derivative 
$\partial_{\bf x} \bar w = \nabla \bar w + \I e [{\bf A}, w]$ can here
also be written as 
$\partial_{\bf x}  =
\nabla -\I e {\bf A}_{t_1}({\bf x}) + \I e {\bf A}_{t_2} ({\bf x} )$.
The $\langle \omega \bar \omega \rangle $ correlations 
solve the differential equation
\begin{eqnarray}
\lefteqn{ \left( 
-{\partial \over \partial t_1} - { \partial \over \partial t_2}
 + D \partial_{\bf x}^2
\right)
\langle w_{t_2 t_1}({\bf x}) \bar w_{t_3 t_4}({\bf x}')  \rangle
}&& \nonumber \\
&&={2\over \pi N_0 }
\delta ({\bf x} -{\bf x}') \delta(t_1-t_3) \delta(t_2-t_4)
.\end{eqnarray}
After introducing the relative times $\eta = t_2 -t_1$,
$\eta' = t_4 -t_3$
and the center-of-mass times,
$t = (t_1+ t_2)/2$, $t' = (t_3 + t_4)/2$, we 
identify this correlator with the diffuson,
\begin{equation}
\langle w_{t_2 t_1} ({\bf x}) 
\bar w_{t_3 t_4}({\bf x'} )\rangle
= -{2 \tau \over \pi N_0 }
 D^{\eta}_{ t t'  } ({\bf x}, {\bf x}')
 \delta ( \eta - \eta' )
,\end{equation}
as it may be seen by comparing with Eq.(\ref{eq31}). 
Notice that the relative time $\eta$ is conserved during the propagation so
that $\eta=\eta'$.
The 
$\langle \bar  w w \rangle$ correlator is the advanced 
counterpart of the
diffuson.
$\langle \bar w \bar w \rangle = 0  $ since there is no term proportional to
$ w w $ in the action.
Finally the $\langle  w w \rangle $ correlator
is given by
\begin{eqnarray}
\langle w_{t_2 t_1} w_{t_3 t_4} \rangle &= & {-\pi N_0} 
\int \D t_1' \dots \D t_4'  
\langle w_{t_2 t_1} \bar w_{t_1' t_2'} \rangle
\langle \bar w_{t_4' t_3'} w_{t_3 t_4} \rangle \nonumber\\
&& \times \partial_{\bf x} F_{t_2' t_4' } \partial_{\bf x} F_{t_3'  t_1'}. 
\end{eqnarray}
This correlator is zero in equilibrium, when
$\partial_{\bf x} F=0 $.
Electron interactions modify these propagators.
By expanding in (\ref{eqA5}) each $Q$-field to first order in $W$,
the interacting part of the action becomes
\begin{equation}
 = -\I (\pi N_0 )^2  {1\over 2} V \int \D t
 ( \bar w F -   F \bar w  )_{tt}
 ( \bar w - w - F \bar w F)_{tt} 
,\end{equation}
which  leads to non-trivial modifications of all the 
$\langle w w \rangle $ correlators. 
Only $\langle \bar \omega \bar \omega \rangle $ is not modified by the
interaction and remains zero.
We consider now 
$\langle w \bar w \rangle $, which in the absence of 
interaction is just the diffuson. We find
\begin{eqnarray}\label{eqA20}
\lefteqn{\left( -{\partial \over \partial t_1} -{\partial \over \partial t_2} + D \partial_{\bf x}^2
\right)
\langle w_{t_2 t_1} \bar w_{t_3 t_4}\rangle_\Phi }&& \cr
& +& \I \pi N_0 V F_{t_2 t_1} \left(
\langle  w_{t_2 t_2}\bar w_{t_3 t_4} \rangle_\Phi 
-\langle w_{t_1 t_1}\bar w_{t_3 t_4} \rangle_\Phi \right) \cr
&= &{2\over \pi N_0} \delta( {\bf x} - {\bf x}') \delta( t_1 -t_3) \delta(t_2- t_4) 
.\end{eqnarray}
To understand the meaning of the above equation,
let us consider first  the limit $t_2 \to t_1$. 
The interaction dependent term does not drop from (\ref{eqA20}) 
since the distribution function is singular in this limit,
$F_{t_2 t_1} \approx -\I /\pi (t_2 - t_1)$.
Upon using this relation we arrive at 
the following simple diffusion equation
\begin{eqnarray}
\lefteqn{\left[ -(1- N_0 V){\partial \over \partial t } + D \partial_{\bf x}^2
\right] \langle w_{tt} \bar w_{t_3 t_4} \rangle_\Phi }
&&\cr
&=& {2\over \pi N_0} \delta( {\bf x} - {\bf x}') \delta(t -t_3) \delta(t - t_4)
.\end{eqnarray}
To make  contact with the diagrammatic calculation,
we identify this correlation function 
with the product of the retarded interaction and diffusion
propagators which have been   defined in Eq.(\ref{eqVRD})
\begin{equation}
-{\pi \over 2} {N_0 V  \over 2}  \langle w_{tt} \bar w_{t_3 t_4} \rangle_\Phi
 = (V^R D)_{tt_3} \delta(t_3 - t_4). 
\end{equation}
The propagator with four different time arguments is finally given by
\begin{eqnarray} \label{eq92}
\lefteqn{ \langle w_{t_2 t_1} \bar w_{t_3 t_4} \rangle_\Phi =
\langle w_{t_2 t_1} \bar w_{t_3 t_4} \rangle  
- \I \pi N_0 V \int \D {\bf x}' \D t_3' \D t_4'
\Large[ } && \cr 
&&
\times
\langle w_{t_2 t_1} \bar w_{t_3' t_4'} \rangle 
\left.
F_{t_4' t_3'}\left(
\langle w_{t_4' t_4'} \bar w_{t_3 t_4} \rangle_\Phi -
\langle w_{t_3' t_3'} \bar w_{t_3 t_4} \rangle_\Phi  \right) \right]
.\end{eqnarray}

\subsection{Correction to the current}
Due to the fluctuations of $Q$ there are corrections to the charge and
current density. Here we calculate these corrections 
by taking into account the Gaussian fluctuations. 
In the derivation, we parallel  the lines of the diagrammatic approach.
We begin by separating the correction to the current in two contributions
$\delta {\bf j} = \delta {\bf j}_a + \delta {\bf j}_b $,
where $\delta {\bf j}_a $ is related to the gradient of the charge 
density. We then proceed by expressing $\delta {\bf j}_b $ in terms of the fields
$\bar \omega$ and $\omega$. In the third step we will then explicitly
include the interaction and  obtain the current formula.
By writing  $Q= Q^{\rm sp} + \delta Q$,  the correction to 
the current reads 
\begin{eqnarray}
\delta {\bf j}({\bf x}, t ) &= &
-{ e \pi D N_0  \over 2}
\langle \Tr \gamma^2 \left\{
  Q^{\rm sp} \partial_{\bf x} \delta Q  
 +\delta Q   \partial_{\bf x} Q^{\rm sp} 
\right.
\nonumber\\
&& \left.
+\delta Q   \partial_{\bf x} \delta Q - \cdots \right\} 
\rangle
.\end{eqnarray}
The dots correspond to the terms
which appear due to $ (\partial_{\bf x} Q) Q$. 
 The term $j_a$ is proportional
to the gradient of the charge density
\begin{eqnarray}
\delta {\bf j}_a({\bf x}, t) &= &- e \pi D N_0 \nabla 
\langle \delta Q^{12}_{tt}({\bf x}) \rangle \\
 &=& -D\nabla \delta \rho ({\bf x}, t)
\end{eqnarray}
and we do not calculate explicitly here.
The second term, $\delta {\bf j}_b$, contains all other contributions. 
By collecting the various summands, one gets
\begin{eqnarray}
\delta {\bf j}_b &=& 
  2F \partial_{\bf x}   \langle \delta  Q^{22} \rangle
+ \langle \delta Q^{11} \rangle \partial_{\bf x} (2F)  \nonumber \\
&&+\langle  \delta Q^{11} \partial_{\bf x}  \delta Q^{12} \rangle
+\langle \delta Q^{12}  \partial_{\bf x} \delta Q^{22} \rangle - \cdots
.\end{eqnarray}
Now we expand $Q$ in powers of $W$ 
and we arrive at
\begin{eqnarray} 
\delta  {\bf j}_{b}  &= & e \pi D N_0 
\left[
F \langle   (\partial_{\bf x} \bar w) w \rangle_{\Phi} +
\langle w (\partial_{\bf x} \bar w) \rangle_{\Phi} F \right] 
\\ 
\label{eq99}
&=& 2\pi e D N_0 \Re \left(\langle w \partial_{\bf x} \bar w \rangle_\Phi
F \right) 
.\end{eqnarray}
In the absence of interactions $\delta {\bf j}_b$  vanishes,
so that taking the interactions into account is essential.
Inserting $\langle w \bar w \rangle_\Phi$ from eq.(\ref{eq92}) 
we find
\begin{eqnarray}
\delta {\bf j}_b & =& 
4 \pi e D N_0 \tau^2 \int \D\eta \D x_1 \D x_2 \Re 
\left\{F^{\vphantom R}_{t-\eta, t }({\bf x})  \right. \nonumber \\
&& 
\times D^\eta_{t-\eta/2, t_1- \eta/2}({\bf x}, {\bf x}_1) 
F_{t_1, t_1 -\eta}({\bf x}_1 )
\nonumber \\
&& \left. \times
V^R_{t_1 t_2}({\bf x}_1, {\bf x}_2 )
( -\I \nabla_{\bf x}) D^0_{t_2, t-\eta}({\bf x}_2, {\bf x} ) 
\right\}
,\end{eqnarray}
which is identical to what we obtained in Eq.(\ref{currentformula}).

Notice that the long derivative in the current formula (\ref{eq99}) 
reduces to the gradient in the final result. This happens since the two
time indices in the relevant $\bar\omega$ field are always equal, 
$\langle w_{tt} \bar w_{t_3 t_4}\rangle_\Phi \propto \delta(t_3 - t_4)$.

\section{Spin-Triplet channel}
\label{sectionTriplet}
Until now we have neglected the spin effects. These arise from the fact that
the diffusion is described by the particle-hole propagator, which may occurs in four
spin states depending on the relative spin of the particle and hole.
In the case of Coulomb long range forces, the most singular contribution comes
from the singlet channel that we have discussed throughout the paper.
In addition to the singlet channel, there are three triplet channels  which
also contribute to the quantum corrections to the current.
In this appendix we extend  our earlier work of Ref.\cite{leadbeater00} to
non-equilibrium. 

In order to take into account the spin-dependent interactions, 
we follow Refs. \cite{finkelstein83,chamon99} and 
start from an interaction which is local in space and time.
The fermionic action is of the type
\begin{equation} \label{eq101}
\I S_{ee} = 
- {\I \over 2 N_0} {\rm Tr} 
\left\{
\Gamma \bar \Psi_s \bar \Psi_{s'} \Psi_{s'} \Psi_s 
- \Gamma_2 \bar \Psi_s \bar \Psi_{s'} \Psi_s \Psi_{s'}
\right\}
,\end{equation}
where $\Gamma$ and $\Gamma_2$ are dimensionless static scattering amplitudes,
and
$\Psi_s$ is an operator for a Fermion with spin $s$.
The trace includes integration  over space and  time contour, as well as  
summation over spin.
The interaction can be written in terms of the charge and spin densities,
\begin{equation}\label{eq102}
\I S_{ee} = - {\I \over 2 N_0} \Tr 
\left\{ \left( \Gamma - {\Gamma_2 \over 2 } \right) \rho \rho 
-{\Gamma_2 \over 2} {\bf s}\cdot {\bf s} \right\}
,\end{equation}
with $\rho = \sum_{s} \bar \Psi_s \Psi_s$ and 
$s^i = \sum_{ss'} \bar \Psi_s 
\sigma^i_{s s'} \Psi_{s'}$.
At this point it becomes convenient to introduce the interaction amplitudes
in the charge (singlet) channel, $\Gamma^s = \Gamma - \Gamma_2/2$ and
in the spin (triplet) channel, $\Gamma^t = -\Gamma_2/2 $.
Then we decouple the interaction with the fields $( \Phi, {\bf B})$ for
the charge and spin. 
By going through the steps of the derivation of the $\sigma$-model one finds
the following modifications of the action
\begin{eqnarray} \label{eq103}
\I S_0 &\to & -{ \pi N_0 \over 4}
\left[ D \Tr \partial_{\bf x} Q_{s s'} \partial_{\bf x} Q_{s's } + 
4 \I \Tr Z \epsilon Q_{ss} \right]\\
\I S_1 &\to & -\I \pi N_0 \Tr
\left[( \Phi_\alpha \delta_{ss'} + {\bf B}_\alpha \cdot {\boldmath \sigma}_{ss'} )
\gamma^\alpha Q_{s' s}
\right]
,\end{eqnarray}
with 
\begin{eqnarray}
- \I \langle \phi_\alpha \phi_\beta  \rangle &=& {1\over 2} 
  {\Gamma_s  \over N_0 } \sigma^x_{\alpha \beta} \\ 
- \I \langle B^i_\alpha B^j_\beta \rangle    &=&
{1\over 2} {\Gamma_t \over N_0 } \sigma^x_{\alpha \beta} 
\delta_{i j}.
\end{eqnarray}
In (\ref{eq103}) we introduced the factor $Z$ which arises in the renormalization 
of the $\sigma$-model\cite{finkelstein83}. One observes that $Z$ can be absorbed 
in a redefinition of the  interaction amplitudes, the quasi-particle diffusion constant, and
 the quasi-particle density of states according to
\begin{eqnarray}
\Gamma^{s,t} &\to & \gamma^{s,t}  =  \Gamma^{s,t}/Z \\
D   & \to & D_{\rm qp}   = D/Z \\
N_0 & \to & N_{\rm qp} = N_0 Z
\end{eqnarray}
We are now ready to consider  the fluctuations. 
In analogy to what has been done in the spinless case, 
 we introduce the charge and spin components of the
fields $w$ and  $\bar w$,
\begin{eqnarray}
w_{ss'}& = &{1\over \sqrt{2}} \sum_{i=0}^3 w^i \sigma^i_{ss'} \\
w^i    & = &{1\over \sqrt{2}} \sum_{ss'} w_{ss'} \sigma^i_{ss'} 
\end{eqnarray}
where $\sigma^0_{ss'}= \delta_{ss'}$ and $\sigma^i_{ss'}$ for $i=1,2,3$
are the usual Pauli matrices. 
In the quadratic fluctuations of the non-interacting action, $S_0$,
the spin and charge terms decouple, since the structure is of the type
$\sum \bar w_{ss'}( \dots  )  w_{s's}=  \sum_i \bar w^i( \dots ) w^i $.
The coupling term $\I S_1$ is to first order in $W$ given by
\begin{equation}
\I S_1 =  -\I \pi N_0 \sqrt{2} \sum_i \Tr \left\{
B^i_\alpha \gamma^\alpha u 
\left( \begin{array}{cc}
0         & w^i \\
-\bar w^i & 0   \end{array}\right) u \right\}
+ \cdots \end{equation}
where for brevity we denoted the scalar field $\Phi$ by $B^0$.
We assume that the saddle point, i.e. the distribution function
and therefore the matrix $u$, does not depend on spin.
The interaction field, $\Phi,  $ is easily integrated out, with the result that at the 
level of the quadratic fluctuations, the  spin and
charge contributions are decoupled even in the presence of the interaction.

The correction to the current is finally found as
\begin{eqnarray}
\delta {\bf j }_b &= &e \pi D N_0 
\sum_{i=0}^3 \Re \langle w^i \nabla_{\bf x}\bar w^i \rangle_{\Phi}^{\rule{0em}{2ex}} F \\
&=& 4 \pi e D_{\rm qp} \tau^2 \sum_{i=0}^3 \int \D \eta \D x_1 \D x_2 
\Re\left\{ 
F_{t-\eta, t}^{\vphantom{t}}({\bf x}) \right.\nonumber \\
&&\times D^{i, \eta}_{t-\eta/2, t_1 -\eta/2}({\bf x}, {\bf x}_1 )
F_{t_1, t_1 -\eta}({\bf x}_1) \nonumber \\
&&\times \left.
\tilde \gamma^{i}_{t_1 t_2}({\bf x}_1, {\bf x}_2 )
(-\I \nabla_{\bf x})D^{i, \eta' =0}_{t_2, t-\eta}({\bf x}_2, {\bf x})
\right\}
,\end{eqnarray}
where again $i=0$ corresponds to the charge and $i=1,2,3$ correspond to spin
channels.
$D^{i,\eta}_{t t'}({\bf x},{\bf x}')$ is the quasi-particle diffusion propagator
in the relevant spin or charge channel,
which obeys the differential equation given in eq.(\ref{eq31}), with the only difference
that the diffusion constant $D$ is replaced by the quasi-particle diffusion
constant $D_{\rm qp}$.
$\tilde \gamma^i$ is the dynamically screened interaction,
\begin{eqnarray}
\tilde \gamma^i({\bf q}, \omega ) & = & \gamma^i [ 1+ \gamma^i \Pi^d/N_{\rm qp} ]^{-1}
\\
\label{eq115}
 & = & \gamma^i 
{ -\I \omega + D_{\rm qp} q^2 \over -\I (1- 2\gamma^i ) \omega +  D_{\rm qp} q^2 }.
\end{eqnarray}
For the explicit calculations it is convenient to consider the product of the
dynamically screened interaction and the retarded diffuson,
\begin{equation}
(\tilde \gamma^i D^i)_{tt'}({\bf x}, {\bf x}' ) =
\int \D x_1 \tilde \gamma^i_{t t_1}({\bf x}, {\bf x}_1 )
D^{i,\eta =0}_{t_1, t'}({\bf x}_1, {\bf x}' ) 
\end{equation}
which solves the diffusion equation
\begin{equation}
\left[ (1 - 2 \gamma^i )\partial_t - D \nabla_{\bf x}^2 \right]
( \tilde \gamma^i D^i)_{tt'}({\bf x}, {\bf x}' ) = {\gamma^i \over \tau}
\delta({\bf x} - {\bf x'}) \delta (t - t').
\end{equation}

\end{appendix}
}
\end{multicols} 

\begin{references}
\bibitem{lee85} P.A. Lee and T.V. Ramakrishnan,
                Rev. Mod. Phys. {\bf 57}, 287 (1985);
		D. Belitz and T.R. Kirkpatrick,
		Rev. Mod. Phys. {\bf 66}, 261 (1994).
\bibitem{altshuler85}B.L. Altshuler and A.G. Aronov, 
                in {\em Electron-Electron Interactions in Disordered Systems},
                edited by M. Pollak and A.L. Efros (North-Holland, 
                Amsterdam, 1985), p. 1.
\bibitem{bergmann84}G. Bergmann, 
		Phys. Rep. {\bf 107}, 1 (1984).
		S. Chakravarty and A. Schmid, 
		Phys. Rep. {\bf 140}, 193 (1986).
\bibitem{gorkov79}L.P. Gorkov, A.I. Larkin, and D.E. Khmel'nitskii
                  Pis'ma Zh. Eksp. Teor. {\bf 30}, 248 (1979) 
		  [JETP Lett. {\bf 30}, 228 (1979)].
\bibitem{abrahams80}E. Abrahams and T.V. Ramakrishnan,
                    J. Non-Cryst. Sol. {\bf 35}, 15 (1980).
\bibitem{altshuler80}B.L. Altshuler, A.G. Aronov, and P.A. Lee,
                Phys. Rev. Lett. {\bf 44}, 1288 (1980);
		B.L. Altshuler, D. Khmel'nitzkii, A.I. Larkin, and P.A. Lee,
		Phys. Rev. B {\bf 22}, 5141 (1980).
\bibitem{larkin86}A.I. Larkin, and D.E. Khmel'nitskii,
		Sov. Phys. JETP {\bf 64}, 1075 (1986).
\bibitem{falko89}V.I. Fal'ko and D.E. Khmel'nitskii,
		Sov. Phys. JETP {\bf 68}, 186 (1989).			
\bibitem{kravtsov93}V.E. Kravtsov and V.I. Yudson, 
                Phys. Rev. Lett. {\bf 70}, 210 (1993).
\bibitem{quantumdot} For a review on the experimental development
                     see L. P. Kouwenhoven et al.  in "
		     Proceeding of the NATO Adavnced Study Institute on
		     Mesoscopic Electron
		     Transport" edited by L.L. Sohn, L.P. Kouwenhoven,
		     and G. Schon, Kluwer Academy Publishers, Dordrecht, 1997,
		     pag.105; see also L.P. Kouwenhoven and C.M. Marcus
		     Physics World, pp35-39 june (1998); 
		     L.P. Kouwenhoven and L. Glazman 
		     Physics World, pp33-38 january (2001)
		     and references therein.
\bibitem{blanter00}Ya.M. Blanter and M. B\"uttiker,
                Phys. Rep. {\bf 336}, 1 (2000).
\bibitem{altshuler98}B.L. Altshuler, M.E. Gershenson, and I.L. Aleiner,
	        Physica E {\bf 3}, 58 (1998).
\bibitem{mohanty97}P. Mohanty, E.M.Q. Jariwala, and R.A. Webb,
                Phys. Rev. Lett. {\bf 78}, 3366 (1997).
\bibitem{mohanty98}P. Mohanty, E.M.Q. Jariwala, and R.A. Webb, 
                Fortschr. Phys. {\bf 46}, 779 (1998).
\bibitem{bergmann90}G. Bergmann, Wei Wei, Yao Zou, and R.M. Mueller, 
                Phys. Rev. B {\bf 41}, 7386 (1990).
\bibitem{giordano91}J. Liu and N. Giordano, Phys. Rev. B {\bf 43}, 1385 (1991). 
\bibitem{raimondi99}R. Raimondi, P. Schwab, and C. Castellani,
		Phys. Rev. B {\bf 60}, 5818 (1999).
\bibitem{leadbeater00}
		M. Leadbeater, R. Raimondi, P. Schwab, and C. Castellani,
		Eur. Phys. J. B {\bf 15}, 277 (2000).
\bibitem{pothier97}H. Pothier, S. Gueron, N.O. Birge, D. Esteve, and M.H. Devoret,
                Phys. Rev. Lett. {\bf 79}, 3490 (1997).
\bibitem{gutmann00}D.B. Gutman and Y. Gefen, preprint, cond-mat/cond-mat/0006468;
                cond-mat/0102134.
\bibitem{keldysh64}L.V. Keldysh  Zh. Eksp. Teor. Fiz. {\bf 47}, 1515 (1964)
                   [Sov. Phys. JETP {\bf 20}, 1018 (1964)]. 	
\bibitem{rammer86}J. Rammer and H. Smith,
                Rev. Mod. Phys. {\bf 58}, 323 (1986).
\bibitem{altshuler78}B.L. Altshuler, Sov. Phys. JETP {\bf 48}, 670 (1978).
\bibitem{blanter96} Ya.M. Blanter, Phys. Rev. B {\bf 54}, 12807 (1996).
\bibitem{kamenev99}A. Kamenev and A. Andreev,
		Phys. Rev. B {\bf 60}, 2218 (1999).
\bibitem{kozub95}V.I. Kozub and A.M. Rudin,
                Phys. Rev. B {\bf 52}, 7853 (1995).
\bibitem{naveh98}Y. Naveh, D.V. Averin, and K.K. Likharev,
	        Phys. Rev. B {\bf 58}, 15371 (1998);
	        Y. Naveh, in {\em XVIII Rencontres de Moriond: Quantum Physics at Mesoscopic
	        Scale}, 
		edited by D.C. Glattli and M. Sanquer (Editions Fronti\'ers, France, 1999).
\bibitem{nagaev95}K.E. Nagaev, Phys. Rev. B {\bf 52}, 4740 (1995).
\bibitem{weber00}H.B. Weber, R. H\"aussler, H. v. L\"ohneysen, and J. Kroha
		Phys. Rev. B {\bf 63} 165426 (2001).
\bibitem{finkelstein83}
		A.M. Finkelstein, 
		Sov. Phys. JETP {\bf 57}, 97 (1983);
		Z. Phys. B {\bf 56}, 189 (1984); in
		{\em Electron Liquid in Disordered Conductors},
		edited by I.M. Khalatnikov, Soviet Scientific Reviews
		Vol 14 (Harwood, London, 1990).		    
\bibitem{hershfield86}S. Hershfield and V. Ambegaokar
                Phys. Rev. B {\bf 34}, 2147 (1986). 
\bibitem{strinati89}G. Strinati, C. Castellani, and C. Di Castro
                Phys. Rev. B {\bf 40}, 12237 (1989).
\bibitem{strinati91}G. Strinati, C. Castellani, C. Di Castro, and G. Kotliar,
                Phys. Rev. B {\bf 44}, 6078 (1991).
\bibitem{chamon99}C. Chamon, A.W.W. Ludwig, and C. Nayak,
		Phys. Rev. B {\bf 60}, 2239 (1999).
\bibitem{feigelman99} M.V. Feigel'man, A.I. Larkin, and M.A. Skvortsov,
                Phys. Rev. B {\bf 61}, 12361 (2000). 	    		
\end{references}
\end{document}